\documentclass[journal]{IEEEtran}

\usepackage{amsmath,amsfonts,amssymb}
\usepackage{newtxtext,newtxmath}
\usepackage{algorithm}
\usepackage{algorithmic}
\usepackage{array}
\usepackage{booktabs}
\usepackage{multirow}
\usepackage{graphicx}
\usepackage{xcolor}
\usepackage{tikz}
\usetikzlibrary{positioning,arrows.meta,shapes.geometric,calc,fit,backgrounds,trees}
\usepackage{tabularx}
\usepackage{url}
\usepackage{cite}
\usepackage{stfloats}
\usepackage{balance}
\usepackage[hidelinks]{hyperref}

\newcolumntype{L}{>{\raggedright\arraybackslash}X}
\newcolumntype{C}{>{\centering\arraybackslash}X}

\newcommand{\lcc}{\mathrm{LCC}}

\begin{document}

\title{Network Topology Evaluation and Dismantling:\\A Taxonomy, Tutorial, and Reproducible Benchmark}

\author{Yizhong~Hu, Yue~Hu, Ao~Shen, Junxiang~Jia, Haoyang~Meng, and~Jianfeng~Guan%
\thanks{The authors are with the School of Computer Science and Technology, \textbf{State Key Laboratory of Networking and Switching Technology}, Beijing
University of Posts and Telecommunications, Beijing 100876, China (e-mail:
\texttt{yzhu@bupt.edu.cn}; \texttt{huyue2023@bupt.edu.cn}; \texttt{shenao0128@bupt.edu.cn};
\texttt{jiajunxiang@bupt.edu.cn}; \texttt{menghy@bupt.edu.cn};
\texttt{jfguan@bupt.edu.cn}). \textit{(Corresponding author: Jianfeng Guan.)}}%
\thanks{Manuscript prepared for arXiv. The benchmark harness and all figures of
this paper are reproducible with the companion code repository
\texttt{https://github.com/kuaiii/dismantling\_topology}.}}

\markboth{Preprint submitted to arXiv}%
{Y.~Hu \MakeLowercase{\textit{et al.}}: Network Topology Evaluation and Dismantling}

\maketitle

\begin{abstract}
Two questions dominate the structural analysis of communication and infrastructure
networks. The first is diagnostic: \emph{how should the quality of a topology be
measured}? The second is adversarial: \emph{how quickly can the topology be broken},
and therefore how much structural margin does it really have? The literature answers
both questions with a large and fragmented body of work---percolation thresholds,
spectral gaps, robustness curves, centrality heuristics, message-passing decimation,
spectral partitioning, and, more recently, graph neural networks and reinforcement
learning---scattered across statistical physics, operations research, and networking,
and seldom compared under a single protocol.
This paper provides a self-contained treatment of the two questions and nothing else:
we deliberately exclude topology \emph{redesign} and recovery-oriented reconfiguration,
and concentrate on (i) the metrics used to evaluate a topology and (ii) the algorithms
used to dismantle it.
We first give a taxonomy of structural evaluation metrics organised by what they
observe---connectivity, distance, spectrum, removal trajectory, and node-level
descriptors---and state for each its definition, input requirement, computational
complexity, and failure mode. We then give a taxonomy of network dismantling methods
covering static centrality heuristics, percolation and collective-influence approaches,
core and decycling methods, spectral partitioning, information-theoretic entanglement,
learning-based policies, and automated heuristic discovery, with an explicit account of
the objective each method actually optimises.
On this basis we present a reproducible benchmark of a unified, open-source library that
exposes twenty-two dismantling methods behind a single interface
\texttt{dismantle(G, method, stop\_condition)} on \texttt{networkx} graphs. Using
$3$ synthetic ensembles (BA, ER, WS) at $N\in\{100,\dots,800\}$, eleven real backbone
topologies, and $10$ random realisations per configuration, we quantify dismantling cost,
trajectory area, secondary-component behaviour, and wall-clock cost, and we test which
topology descriptors actually predict dismantling difficulty.
Finally, we give a step-by-step tutorial that turns the taxonomy into an auditable
workflow---question, graph, protocol, observables, verification, interpretation---and a
checklist of nine recurring interpretation errors.
The main empirical lessons are that (i) trajectory area and threshold cost agree in
aggregate but invert specific method pairs, so the summary statistic must be named;
(ii) the advantage of targeted over random removal tracks degree heterogeneity and grows
with $N$ on heterogeneous ensembles while staying small on small-world ones;
(iii) collective-influence and spectral methods are the consistent winners, whereas
eigenvector centrality is no better than random removal and betweenness is dominated on
both effectiveness and cost; and (iv) pooled descriptor correlations drastically
overstate predictive power, because degree heterogeneity explains $95\%$ of the variance
across ensembles but only $25\%$ within one.
\end{abstract}

\begin{IEEEkeywords}
Network topology, robustness metrics, percolation, network dismantling,
structural vulnerability, benchmark, tutorial, reproducibility.
\end{IEEEkeywords}

\section{Introduction}
\label{sec:intro}

\IEEEPARstart{A}{communication} network is, before it is a protocol stack, a graph.
Routers, satellites, base stations, and edge devices become vertices; fibre, radio,
free-space optical, and acoustic links become edges. Whatever the network is asked to
do---carry traffic, disseminate alerts, synchronise clocks---it can only do it through
paths that exist in that graph. It follows that the \emph{shape} of the graph is a
first-order engineering asset, and that two questions about that shape recur in every
phase of the lifecycle, from design review to post-incident forensics.

The first question is diagnostic: \emph{how should the quality of a topology be
measured}? The second is adversarial: \emph{how quickly can this topology be broken},
and hence how much structural margin does it actually possess? The two questions are
not independent. A metric such as ``the network is 92\% connected'' is meaningless
until one specifies \emph{under which disturbance} the 92\% is attained. Conversely, a
dismantling algorithm is not merely an attack recipe; run to completion it generates a
\emph{curve}, and that curve is the most informative single object one can compute about
the structural margin of a topology.

\subsection{Why the Two Questions Must Be Treated Together}

Consider the classical robustness picture. Under random edge or vertex loss, a
scale-free network is remarkably resilient, because the probability that an
undirected random strike hits one of the few hubs is small \cite{albert2000error,
cohen2000resilience}. Under \emph{targeted} loss of the same hubs the picture inverts,
and a handful of removals can fragment the network \cite{cohen2001breakdown}. The
metric used in the first statement (the size of the largest connected component after a
fraction $q$ of random failures) and the metric used in the second (the same quantity
after $q$ removals chosen by degree) are formally identical. They differ only in the
disturbance. This is the single most important observation for anyone evaluating a
topology: \emph{a structural metric is a functional of a disturbance protocol}, and
reporting one without the other is not a measurement but a convention.

Network dismantling makes this observation operational. The dismantling problem asks
for the smallest set of vertices whose removal reduces the largest connected component
below a prescribed target, and it is NP-hard \cite{Braunstein2016NetworkD}. Because the
exact optimum is out of reach, the field has produced a sequence of approximations:
greedy centrality heuristics, percolation-based decimation with collective influence
\cite{Morone2015InfluenceMI}, core decomposition and decycling
\cite{Zdeborov2016FastAS}, spectral bisection with vertex cover
\cite{Braunstein2016NetworkD, Ren2018GeneralizedND}, belief-propagation decimation
\cite{Mugisha2016IdentifyingOT}, information-theoretic entanglement
\cite{Huang2024IdentifyingKP}, and, in the last five years, learned policies
\cite{Fan2020FindingKP, Grassia2021MachineLD, Gu2025DeeplearningaidedDO}. Each of these
methods is, for evaluation purposes, a \emph{lower bound generator}: it produces an
achievable removal cost, and therefore an upper bound on how much structural margin the
topology really has against an intelligent adversary.

\subsection{Scope and Deliberate Exclusions}

The literature that touches these questions is enormous, and it is easy for a survey to
expand until it becomes a survey of network science. We therefore fix a narrow scope and
state it explicitly.

\begin{itemize}
\item \textbf{Included.} Structural evaluation metrics for a \emph{given, static,
unweighted or weighted} graph under vertex or edge removal; the percolation theory that
gives these metrics their meaning; algorithms that compute removal sets or removal
orders; and the empirical protocol needed to compare such algorithms fairly.
\item \textbf{Excluded.} Topology \emph{redesign}: rewiring, edge addition, onion-like
construction, and the optimisation literature that seeks to \emph{improve} a topology
\cite{herrmannonionlike2011, Zhou2017ATM, Buesser2011OptimizingTR}. Also excluded are
recovery-oriented reconfiguration and restoration scheduling
\cite{oneil2025enhancingcritical}, traffic-engineering metrics that require a demand
matrix, and cascading-failure dynamics, which we treat only as a boundary condition
(Section~\ref{sec:eval}) because they add a load-redistribution layer on top of the
purely structural problem.
\end{itemize}

This exclusion is not arbitrary. Redesign and evaluation answer different questions and
use different evidence: redesign papers need an optimisation baseline and report
improvement over it, whereas evaluation papers need a disturbance protocol and report a
curve. Mixing them produces surveys in which a metric cannot be traced to the
disturbance that gives it meaning.

\subsection{Application Contexts}

The abstract questions above acquire urgency in four settings that recur throughout this
paper. \emph{Space and airborne networks}: a LEO constellation reconfigures its
inter-satellite links on a tens-of-seconds timescale, so the topology is a sequence of
snapshots rather than a fixed object, and a structural evaluation must be indexed by the
snapshot it refers to \cite{chen2022low, farhan2025rdg, kang2025deepdive,
digitaltwinleo2026}. \emph{Constrained and remote networks}: underwater acoustic and
industrial deployments have low degree, long links, and repair times measured in hours,
which removes redundancy as an option and makes the \emph{cheapest} dismantling set the
relevant vulnerability \cite{akyildiz2005uans, sarker2024low}. \emph{Cyber-physical and
IoT deployments}: heterogeneity and scale make per-device analysis impossible, so the
question becomes which aggregate metric is both computable and informative
\cite{berger2021survey, kim2022survey, alhidaifi2024survey}. \emph{Interdependent
infrastructure}: when power, communication, and transport couple, a structural failure in
one layer propagates through the coupling, and the dismantling question becomes a
multi-layer one \cite{helbing2013globally, kivela2014multilayer, Gao2011NetworksFF}.

In each case the underlying object is the same: a graph, a disturbance, and a trajectory.
The classical apparatus for describing such graphs---random-graph evolution
\cite{erdos1960evolution}, small-world structure \cite{watts1998collective}, preferential
attachment \cite{barabasi1999emergence}, and the standard textbook treatments
\cite{newman2010networks, barabasi2016networkscience}---supplies the vocabulary, and the
percolation literature supplies the quantitative predictions that make descriptors
useful before any simulation is run.

\subsection{Positioning with Respect to Existing Surveys}

Several excellent references already cover parts of this ground, and we do not attempt
to replace them. Reference \cite{lou2023structural} catalogues \emph{a posteriori}
robustness measures and their algebraic properties; \cite{wan2021survey} surveys
centrality metrics with an eye to resilience analysis; \cite{Li2021PercolationOC} is the
standard reference for percolation on networks; \cite{Wandelt2018ACA} compares
dismantling approaches empirically; and \cite{pedroso2025survey} catalogues faults and
vulnerabilities of topological connectivity from a networking standpoint. Very recent
work has begun to standardise benchmarking itself \cite{netbreaker2026}.

What is missing, and what this paper supplies, is a \emph{single} treatment in which
(i) metrics are indexed by the disturbance they presuppose, (ii) dismantling algorithms
are indexed by the objective they actually optimise and the complexity they incur,
(iii) both are instantiated in one open library whose behaviour can be reproduced, and
(iv) the resulting numbers are turned into a tutorial procedure that a practitioner can
follow on their own topology. The empirical component is not a leaderboard; it is a
demonstration that the choice of summary statistic changes the ranking, and that this
must be reported.

\subsection{Contributions}

\begin{enumerate}
\item \textbf{A disturbance-indexed taxonomy of topology metrics}
(Section~\ref{sec:metrics}). We organise metrics by the object they observe
---connectivity, distance, spectrum, removal trajectory, and node-level descriptors---
and record for each its definition, its input requirement, its asymptotic cost, and the
specific way it can mislead.

\item \textbf{A objective-indexed taxonomy of dismantling algorithms}
(Section~\ref{sec:methods}). We separate methods by what they optimise (a static score,
a percolation threshold, a cycle rank, a spectral cut, an information-theoretic
functional, or a learned policy), because methods with different objectives are not
interchangeable even when they are benchmarked on the same curve.

\item \textbf{A reproducible benchmark of a unified library}
(Sections~\ref{sec:library}--\ref{sec:results}). Twenty-two methods are exposed through
one call, \texttt{dismantle(G, method, stop\_condition)}, on \texttt{networkx} graphs.
We evaluate ten of them across four sizes and fourteen at one size, on three synthetic ensembles
and eleven real backbone topologies with ten realisations per configuration, and we
publish the harness alongside the paper.

\item \textbf{Evidence on which descriptors predict difficulty}
(Section~\ref{sec:results}). We regress dismantling cost against ten topology
descriptors across the synthetic ensemble and show which of them carry predictive
information, and which are redundant.

\item \textbf{A tutorial and an error checklist}
(Section~\ref{sec:tutorial}). The tutorial converts the taxonomy into an ordered
procedure with executable fragments, and closes with nine documented interpretation
errors, each paired with the check that detects it.
\end{enumerate}

\begin{figure*}[t]
\centering
\begin{tikzpicture}[
  font=\footnotesize,
  box/.style={draw, rounded corners=2pt, align=center, minimum height=15mm,
              text width=29mm, inner sep=2pt, fill=gray!6},
  chip/.style={draw, rounded corners=1pt, align=center, inner sep=2pt,
               font=\scriptsize, fill=white, text width=34mm},
  ar/.style={-{Latex[length=2mm]}, thick},
  dar/.style={-{Latex[length=2mm]}, thick, dashed}
]
\node[box] (G) at (0,0)
  {\textbf{Topology}\\$G=(V,E)$\\[1pt]
   {\scriptsize $\langle k\rangle$, $\mathrm{CV}(k)$, $C$, $r$, $\lambda_2$, $k$-core}};
\node[box, right=6mm of G] (D)
  {\textbf{Disturbance}\\protocol $q$\\[1pt]
   {\scriptsize random, degree,\\ CI, core, spectral}};
\node[box, right=6mm of D] (S)
  {\textbf{Trajectory}\\$S(q)=|\lcc|/N$\\[1pt]
   {\scriptsize $q\in[0,1]$, full replay}};
\node[box, right=6mm of S] (M)
  {\textbf{Summary}\\statistics\\[1pt]
   {\scriptsize $q_c(\tau)$, AUC, $R$,\\ $\max S_2$, wall-clock}};

\draw[ar] (G) -- (D);
\draw[ar] (D) -- (S);
\draw[ar] (S) -- (M);

\coordinate (L) at ($(G.south)+(0,-7mm)$);
\coordinate (R) at ($(M.south)+(0,-7mm)$);
\draw[dar] (M.south) -- (R) -- (L) -- (G.south);
\node[font=\scriptsize, align=center, fill=white, inner sep=2pt]
  at ($(L)!0.5!(R)$) {redesign / reconfiguration: \emph{excluded from this paper}};
\end{tikzpicture}
\caption{The evaluation loop assumed throughout this paper. A topology is described by
static descriptors (left), stressed by a declared disturbance protocol (centre), and
observed as a removal trajectory $S(q)$ from which summary statistics are extracted
(right). The decision layer consumes the summaries; the dashed edge marks the feedback
that a negative evaluation would trigger in a design process---which is precisely the
redesign loop that this paper excludes from its scope.}
\label{fig:framework}
\end{figure*}
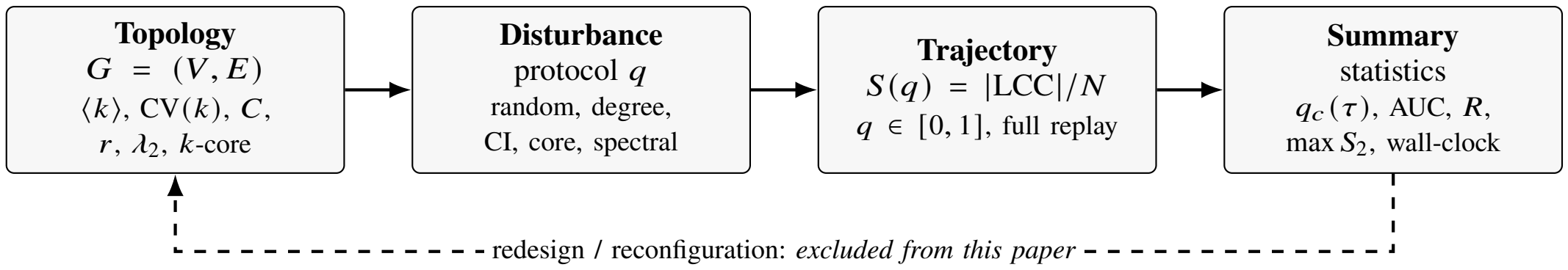

\subsection{Reading Guide and Reproducibility Statement}

Sections~\ref{sec:metrics} and~\ref{sec:methods} are the survey; Sections~\ref{sec:eval}
and~\ref{sec:library} define the measurement protocol and the software;
Sections~\ref{sec:results} and~\ref{sec:tutorial} are the empirical and pedagogical
parts. A reader interested only in ``what should I compute'' can read
Section~\ref{sec:metrics}, then Section~\ref{sec:tutorial}. A reader interested only in
``which algorithm should I use'' can read Section~\ref{sec:methods}, then
Section~\ref{sec:results}.

All numbers reported in Sections~\ref{sec:results} and~\ref{sec:tutorial} were produced
by the harness in \texttt{bench/} against library commit \texttt{fb292f6}. Graphs,
seeds, removal sequences, trajectories, and per-run metrics are retained, and every
table in the paper is generated by a script from those files rather than transcribed.
Where a method could not be executed in our environment for a documented reason
(Section~\ref{sec:library}), it is discussed but not measured, and its reported
performance is attributed to its original publication.

\section{A Disturbance-Indexed Taxonomy of Topology Metrics}
\label{sec:metrics}

\subsection{Notation and Standing Assumptions}

Let $G=(V,E)$ be a simple undirected graph with $N=|V|$ and $M=|E|$, and let
$A\in\{0,1\}^{N\times N}$ be its adjacency matrix with degree matrix $D=\mathrm{diag}(k)$
and Laplacian $L=D-A$. A \emph{removal sequence} is an ordering
$\pi=(v_1,\dots,v_N)$ of $V$; the prefix $\pi_{1:k}$ is the set removed at budget $k$,
and $G_k=G[V\setminus\pi_{1:k}]$ is the residual graph. We write
\begin{equation}
S(k)=\frac{|C_{\max}(G_k)|}{N},\qquad q=\frac{k}{N},
\label{eq:S}
\end{equation}
where $C_{\max}(\cdot)$ denotes the largest connected component (LCC). The
normalisation by the \emph{original} $N$ is deliberate and is discussed in
Section~\ref{sec:eval}: it keeps removed vertices in the reference population, so that a
small surviving fragment cannot appear intact merely because it contains every survivor.

Throughout, ``robustness'' denotes the behaviour of $S$ under a specified disturbance,
and ``resilience'' is reserved for the time-domain notion that additionally involves
recovery \cite{liu2022network, artime2024robustness}. This paper measures robustness in
the first sense; the temporal notion is only invoked in Section~\ref{sec:discussion}.

\subsection{Connectivity Metrics}

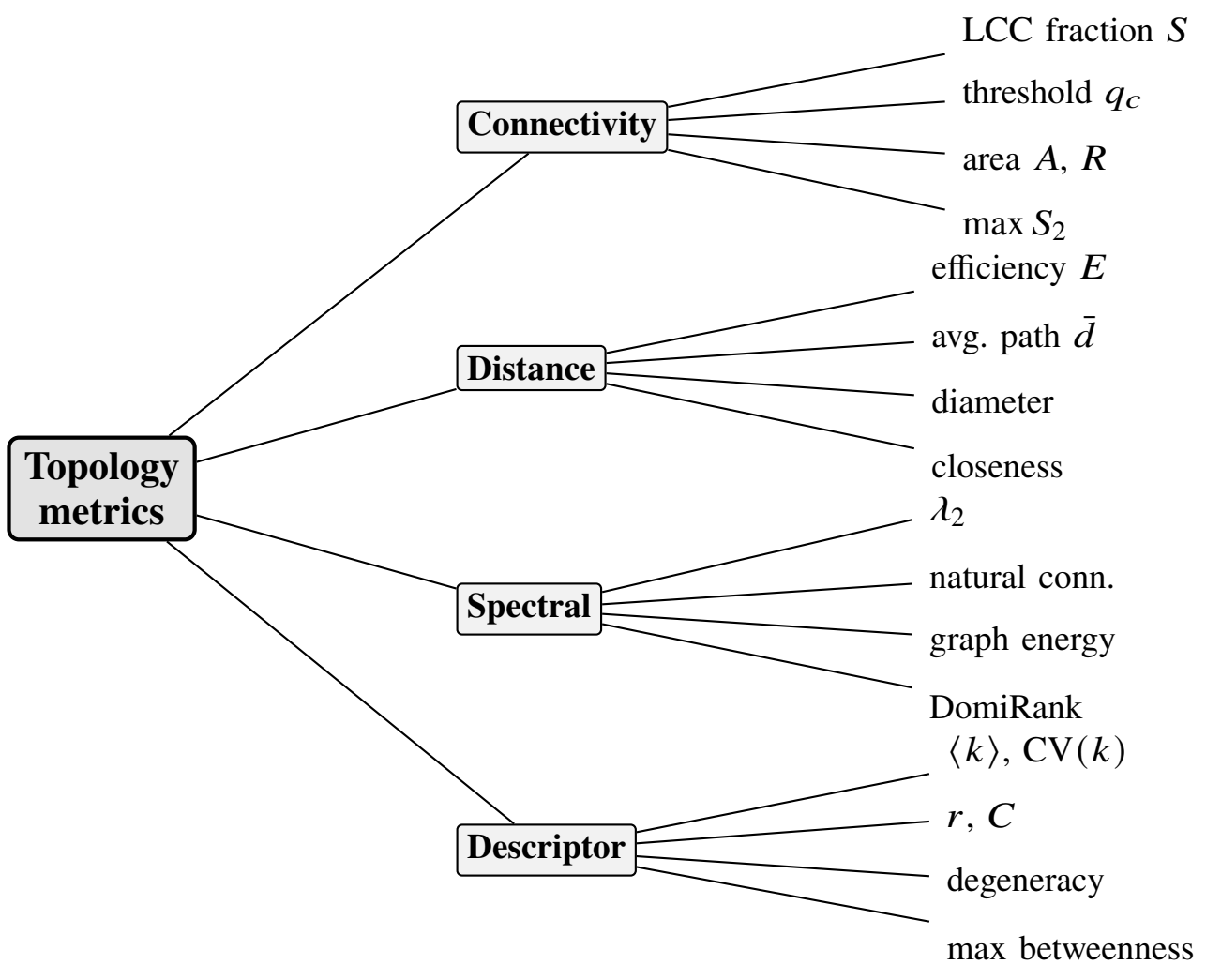
\begin{figure*}[t]
\centering
\begin{tikzpicture}[
  grow=right,
  font=\scriptsize,
  root/.style={draw, thick, rounded corners=2pt, fill=gray!22, inner sep=3pt,
               font=\footnotesize\bfseries, anchor=west, align=center},
  fam/.style={draw, rounded corners=1pt, fill=gray!10, inner sep=2pt,
              font=\scriptsize\bfseries, anchor=west, align=center},
  every node/.style={anchor=west, align=center},
  level 1/.style={level distance=25mm, sibling distance=17mm},
  level 2/.style={level distance=27mm, sibling distance=4.6mm},
  edge from parent/.style={draw, thin}
]
\node[root] {Topology\\metrics}
  child { node[fam] {Descriptor}
    child { node {max betweenness} }
    child { node {degeneracy} }
    child { node {$r$, $C$} }
    child { node {$\langle k\rangle$, $\mathrm{CV}(k)$} }
  }
  child { node[fam] {Spectral}
    child { node {DomiRank} }
    child { node {graph energy} }
    child { node {natural conn.} }
    child { node {$\lambda_2$} }
  }
  child { node[fam] {Distance}
    child { node {closeness} }
    child { node {diameter} }
    child { node {avg.\ path $\bar d$} }
    child { node {efficiency $E$} }
  }
  child { node[fam] {Connectivity}
    child { node {$\max S_2$} }
    child { node {area $A$, $R$} }
    child { node {threshold $q_c$} }
    child { node {$\lcc$ fraction $S$} }
  };
\end{tikzpicture}
\caption{Taxonomy of topology evaluation metrics by observable. The four families are
complementary rather than competing: descriptors are computed before any disturbance,
connectivity and distance metrics observe a disturbed graph, and spectral metrics
interpolate between the two because they degrade smoothly as disconnection approaches.}
\label{fig:taxmetrics}
\end{figure*}

\subsubsection{Largest connected component}
$S(k)$ of \eqref{eq:S} is the canonical observable. Its computation costs
$O(N+M)$ per snapshot with a breadth-first traversal, so a full trajectory costs
$O(N(N+M))$---linear in $N$ snapshots, which is why full trajectories up to
$N\approx10^5$ are practical in compiled code and why the per-step recomputation
dominates the cost of naive evaluation. $S$ is dimensionless, monotone non-increasing
along a fixed sequence, and directly interpretable as the fraction of devices that can
still reach each other.

\subsubsection{Percolation threshold}
For random removal with retention probability $p=1-q$, percolation theory predicts a
critical $p_c$ at which $S$ collapses. For a configuration-model graph with degree
distribution $p_k$,
\begin{equation}
p_c=\frac{\langle k\rangle}{\langle k^2\rangle-\langle k\rangle},
\label{eq:pc}
\end{equation}
so that the removal threshold is $q_c^{\mathrm{rnd}}=1-p_c=1-1/(\kappa-1)$ with
$\kappa=\langle k^2\rangle/\langle k\rangle$ \cite{molloy1995critical,
dorogovtsev2008critical, Li2021PercolationOC}. Equation~\eqref{eq:pc} is the reason
degree heterogeneity is a \emph{first} diagnostic: it fixes how much random loss a
topology tolerates before any simulation is run. For ER graphs $\kappa=\langle
k\rangle+1$ and \eqref{eq:pc} reduces to $p_c=1/\langle k\rangle$; for a pure scale-free
degree sequence with exponent $\gamma\le3$ the second moment diverges and
$p_c\to0$ in the thermodynamic limit---the mathematical content of the statement that
scale-free topologies are robust to random failure and fragile to targeted attack
\cite{albert2000error, cohen2000resilience, cohen2001breakdown}. The mean-field form
\eqref{eq:pc} is exact on configuration-model graphs \cite{callaway2000robustness} and
becomes an approximation in the presence of clustering, degree correlations, or finite
size, where it can be biased by several percent and must be corrected or estimated
numerically \cite{serrano2006percolation, rong2022estimation, rohlf2002criticality}.

\subsubsection{Secondary component and fragmentation}
Write $S_2(k)=|C_2(G_k)|/N$ for the second-largest component. Its maximum over $k$,
$\max_k S_2$, locates the ``shattering'' regime: a large $S_2$ signals a topology that
breaks into a few comparable pieces rather than dissolving into singletons. $S_2$ is
therefore a diagnostic of \emph{how} a network fails, not \emph{how much}; it has no
universal better-or-worse direction and should be read alongside $S$
\cite{Wandelt2018ACA, Schneider2011MitigationOM}.

\subsubsection{Articulation points and $k$-cores}
A vertex whose removal increases the number of components is an articulation point; its
existence bounds the vertex connectivity from above and identifies single points of
failure that no averaged metric reveals \cite{Tian2016ArticulationPI}. The $k$-core is
the maximal subgraph of minimum degree $\ge k$; the degeneracy of $G$ (the largest $k$
with a non-empty $k$-core) measures how much ``core'' a removal procedure must destroy
before the graph becomes a forest. Core-based dismantling exploits precisely this
structure \cite{Zdeborov2016FastAS}, and the core number of a vertex is a cheap proxy
for its structural importance \cite{kitsak2010identification}.

\subsection{Distance and Efficiency Metrics}

Global efficiency \cite{tutorial_latora2001efficiency} averages reciprocal shortest-path
lengths,
\begin{equation}
E_{\mathrm{glob}}(k)=\frac{1}{N(N-1)}\sum_{\substack{i,j\in V\\ i\neq j}}
\frac{1}{d_k(i,j)},
\label{eq:eff}
\end{equation}
with $1/d=0$ when $i$ and $j$ are unreachable or removed. Retaining $N(N-1)$ in the
denominator makes $E_{\mathrm{glob}}$ sensitive to \emph{three} distinct losses at once:
lost endpoints, lost reachability, and lengthened surviving paths. This makes it a better
proxy for service degradation than $S$ in networks where connectivity survives but
latency collapses---for instance when a dismantling procedure leaves a long chain---
but it also means that $E_{\mathrm{glob}}$ and $S$ answer different questions and must
not be used interchangeably. Average shortest-path length and diameter are the
complementary ``how far'' summaries; both are undefined on disconnected graphs unless
restricted to the LCC, and that restriction silently changes the population being
averaged.

\subsection{Spectral Metrics}

\subsubsection{Algebraic connectivity}
The second smallest Laplacian eigenvalue $\lambda_2$ is zero if and only if $G$ is
disconnected, and by Cheeger-type inequalities it bounds how easily the graph can be cut
into two large pieces. It is therefore a \emph{continuous} counterpart of connectivity:
$\lambda_2$ degrades smoothly as the graph approaches disconnection, giving an early
warning that $S$ cannot provide (because $S$ stays near 1 until the graph actually
splits). Its cost is $O(N^3)$ for a dense eigensolver and $O(M\sqrt{N})$--$O(N^2)$ for
sparse iterative methods on the few smallest eigenvalues, which is acceptable for design
review but not for per-step use inside a removal loop.

\subsubsection{Natural connectivity and spectral radius}
Natural connectivity, $\bar\lambda=\ln\left(\frac1N\sum_i e^{\lambda_i}\right)$ over
Laplacian eigenvalues, aggregates all eigenvalues and is monotonically related to the
number of closed walks; the spectral measure of robustness of \cite{Wu2011SpectralMO}
and the graph-energy-based indicators of \cite{edwards2024method, edwards2024impact}
belong to the same family. These measures are smooth, differentiable, and cheap to
evaluate incrementally, which makes them attractive as surrogates inside optimisation
loops---but their absolute values are not comparable across graphs of different size or
density.

\subsubsection{Node-dominance centrality}
DomiRank \cite{engsig2024domirank} assigns each vertex a dominance score through a
linear system that mimics competitive spread, and its \emph{inverse} ranks vertices by
structural fragility. It is notable as a bridge between the metric and the algorithm
halves of this paper: a centrality that predicts which vertex removal hurts most is,
operationally, a one-step dismantling heuristic.

\subsection{Trajectory Metrics}

Because a single snapshot is a poor summary, the literature converged on functionals of
the whole curve \eqref{eq:S}.

\subsubsection{Robustness $R$}
Schneider \emph{et al.} \cite{Schneider2011MitigationOM} define
\begin{equation}
R=\frac{1}{N}\sum_{k=1}^{N}S(k),
\label{eq:R}
\end{equation}
the mean LCC fraction along the sequence. $R$ rewards topologies that retain
connectivity for long, and it is the quantity maximised by onion-like constructions
\cite{herrmannonionlike2011}. It is a rectangle-rule approximation to the area under $S$.

\subsubsection{Trajectory area}
The trapezoidal area
\begin{equation}
A=\int_0^1 S(q)\,\mathrm{d}q
\label{eq:A}
\end{equation}
is the same idea with a finer quadrature and is what our harness reports as ``AUC''.
$A$ and $R$ differ by $O(1/N)$ and are interchangeable in practice; both are
\emph{aggregate} measures that average over the entire trajectory, including the
uninteresting tail where $S\approx0$.

\subsubsection{Threshold cost}
The most decision-relevant summary is the cost of reaching a target,
\begin{equation}
q_c(\tau)=\min\{q: S(q)\le\tau\},\qquad \tau\in\{0.01,0.1\},
\label{eq:qc}
\end{equation}
the operational dismantling threshold. The $1\%$ convention is standard in the
dismantling literature \cite{Braunstein2016NetworkD, Ren2018GeneralizedND} and matches
the theoretical notion of a vanishing giant component up to finite-size effects;
$\tau=0.1$ is a common engineering target. \emph{$q_c$ and $A$ are not equivalent.} Two
procedures can reach the same target at the same cost while differing substantially in
the middle of the curve, and conversely a procedure that is uniformly mediocre in the
middle can reach the target earlier because of a late cascade. Section~\ref{sec:results}
quantifies this disagreement.

\subsection{Node-Level Descriptors}

Descriptors are cheap, single-number summaries of the topology that are used for
pre-screening rather than for evaluation proper: mean degree $\langle k\rangle$, degree
heterogeneity $\mathrm{CV}(k)=\sigma_k/\langle k\rangle$, maximum degree, assortativity
$r$, average clustering $C$, density, average path length, diameter, degeneracy, and
maximum betweenness. Their appeal is cost: all are computable in at most
$O(NM)$---betweenness being the exception---and all can be obtained before any removal
experiment. Their danger is over-interpretation: $\mathrm{CV}(k)$ correlates with
robustness on synthetic ensembles largely because those ensembles differ in
$\kappa$, and the correlation weakens once $\kappa$ is held fixed.
Section~\ref{sec:results} tests this directly.

\begin{table*}[t]
\caption{Taxonomy of topology evaluation metrics. Cost is per evaluation on a graph with
$N$ vertices and $M$ edges; ``snapshot'' refers to a single residual graph $G_k$.
``Blind spot'' states the failure mode that the metric cannot represent.}
\label{tab:metrics}
\centering
\footnotesize
\begin{tabularx}{\textwidth}{@{}l l L l L@{}}
\toprule
\textbf{Family} & \textbf{Metric} & \textbf{Definition} & \textbf{Cost} & \textbf{Blind spot} \\
\midrule
\multirow{4}{*}{Connectivity}
& $S(q)$ & $|C_{\max}(G_k)|/N$ & $O(N{+}M)$/snapshot & Says nothing until the graph splits \\
& $q_c(\tau)$ & first $q$ with $S\le\tau$ & $O(N(N{+}M))$ & Depends on $\tau$; tail-sensitive \\
& $\max S_2$ & peak second component & $O(N(N{+}M))$ & No better/worse direction \\
& deg.\ / $k$-core & degeneracy, core number & $O(N{+}M)$ & Ignores global position \\
\midrule
\multirow{3}{*}{Distance}
& $E_{\mathrm{glob}}$ & \eqref{eq:eff} & $O(N(M{+}N\log N))$ & Conflates loss and detour \\
& $\bar d$, diam. & LCC-restricted averages & $O(NM)$ & Undefined when disconnected \\
& closeness & mean $1/d$ per vertex & $O(NM)$ & Hides worst-case pairs \\
\midrule
\multirow{3}{*}{Spectral}
& $\lambda_2$ & 2nd Laplacian eigenvalue & $O(N^3)$ dense & Continuous: no threshold \\
& $\bar\lambda$ & $\ln\frac1N\sum e^{\lambda_i}$ & $O(N^3)$ & Not comparable across $N$ \\
& DomiRank & linear dominance system & $O(N^3)$ & One-step, not a trajectory \\
\midrule
\multirow{3}{*}{Trajectory}
& $R$ & \eqref{eq:R} & $O(N(N{+}M))$ & Averages over the trivial tail \\
& $A$ & \eqref{eq:A} & $O(N(N{+}M))$ & Ignores where the drop happens \\
& $q_c$ vs.\ $A$ & ranking comparison & --- & Rankings can disagree \\
\midrule
\multirow{3}{*}{Descriptor}
& $\mathrm{CV}(k)$ & $\sigma_k/\langle k\rangle$ & $O(N)$ & Proxies $\kappa$, not structure \\
& $r$ & degree assortativity & $O(M)$ & Sign flips across ensembles \\
& $C$ & avg.\ clustering & $O(N\langle k\rangle^2)$ & Unrelated to cut size \\
\bottomrule
\end{tabularx}
\end{table*}

\subsection{What the Metrics Do Not Measure}

Three limitations recur and are worth isolating, because each corresponds to a documented
error in the applied literature.

\emph{(i) Metrics are relative to a protocol.} $S$, $R$, $A$, and $q_c$ are all
functionals of a removal sequence $\pi$. Reporting ``robustness $=0.42$'' without
specifying whether $\pi$ is random, degree-ordered, or optimised is not reproducible.
This is why the disturbance must be part of the metric definition and why the taxonomy is
indexed by disturbance in the first place.

\emph{(ii) Aggregate metrics hide the mechanism.} A topology can obtain a good $A$ by
being uniformly mediocre, and a bad $A$ by being robust until a sudden collapse. The
mechanism (gradual erosion vs.\ abrupt shattering) is visible only in the shape of
$S(q)$ and in $S_2(q)$, which is why we plot both
(Fig.~\ref{fig:curves}) rather than reporting scalars alone.

\emph{(iii) Different measures rank the same networks differently.} Comparative studies
of robustness measures report rank reversals across measures on identical graph sets
\cite{Liu2017ACS}, and application-specific variants---connection-pattern measures
\cite{cerqueti2019measuring}, quantitative resilience indices
\cite{ahmadian2020quantitative}, transportation-specific topology measures
\cite{zhang2015assessing}---inherit this ambiguity. The practical implication is that a
single number should never be reported as ``the robustness''; the family and the
functional form must be named.

\emph{(iv) Static descriptors are not evaluations.} A descriptor is a property of the
graph; an evaluation is a property of the graph \emph{under a disturbance}. Replacing the
latter by the former is the most common shortcut in practice, and
Section~\ref{sec:results} shows how much explanatory power it actually retains.

\section{An Objective-Indexed Taxonomy of Dismantling Algorithms}
\label{sec:methods}

\subsection{Problem Statement}

Given $G=(V,E)$ and a target fraction $\tau$, the \emph{network dismantling} problem is
\begin{equation}
\min_{S\subseteq V}\ |S|\quad\text{s.t.}\quad |C_{\max}(G[V\setminus S])|\le\tau N .
\label{eq:dismantle}
\end{equation}
The problem is NP-hard \cite{Braunstein2016NetworkD}, and it is closely related to the
minimum feedback vertex set: removing a feedback vertex set destroys all cycles, after
which the residual graph is a forest and can be fragmented cheaply by a tree-breaking
procedure. This two-stage structure---\emph{decycle, then break trees}---organises a
large part of the algorithmic literature and explains why the best-performing methods
tend to combine a global structural step with a local refinement.

Two features of \eqref{eq:dismantle} deserve emphasis because they are frequently
misstated. First, the objective is \emph{not} submodular in $S$: adding a vertex to $S$
can reduce $|C_{\max}|$ by more when $S$ is already large than when it is small, so the
standard $(1-1/e)$ greedy guarantee does not apply, and greedy procedures are heuristics
without worst-case approximation bounds. Second, the problem is stated for a
\emph{worst-case} adversary: a solution to \eqref{eq:dismantle} is a certificate that
the topology can be broken at that cost, not a prediction of what a particular attacker
will do. Evaluation and attack are therefore the same computation viewed from two sides.

\subsection{Static Centrality Heuristics}

\begin{figure*}[t]
\centering
\begin{tikzpicture}[
  grow=right,
  font=\scriptsize,
  root/.style={draw, thick, rounded corners=2pt, fill=gray!22, inner sep=3pt,
               font=\footnotesize\bfseries, anchor=west, align=center},
  fam/.style={draw, rounded corners=1pt, fill=gray!10, inner sep=2pt,
              font=\scriptsize\bfseries, anchor=west, align=center},
  every node/.style={anchor=west, align=center},
  level 1/.style={level distance=26mm, sibling distance=15mm},
  level 2/.style={level distance=27mm, sibling distance=4.2mm},
  edge from parent/.style={draw, thin}
]
\node[root] {Dismantling\\algorithms}
  child { node[fam] {Learning}
    child { node {AutoSND} }
    child { node {NIRM} }
    child { node {FINDER} }
    child { node {GDM} }
  }
  child { node[fam] {Spectral}
    child { node {directed} }
    child { node {embedding} }
    child { node {EGND} }
    child { node {GND} }
  }
  child { node[fam] {Core /\\decycling}
    child { node {tree break} }
    child { node {decycler} }
    child { node {min-sum} }
    child { node {CoreHD} }
  }
  child { node[fam] {Percolation}
    child { node {EI} }
    child { node {BPD (cavity)} }
    child { node {CI + reinsert} }
    child { node {CI$_\ell$} }
  }
  child { node[fam] {Static\\centrality}
    child { node {$k$-core} }
    child { node {eigenvector, PR} }
    child { node {betweenness} }
    child { node {degree} }
  };
\end{tikzpicture}
\caption{Taxonomy of dismantling algorithms by the surrogate they optimise. The five
families are ordered by the amount of global structure they exploit, from a one-step local
score (static centrality) to a learned policy that is itself adaptive. ``AutoSND'' denotes
automated discovery of the heuristic rather than a fixed algorithm.}
\label{fig:taxmethods}
\end{figure*}
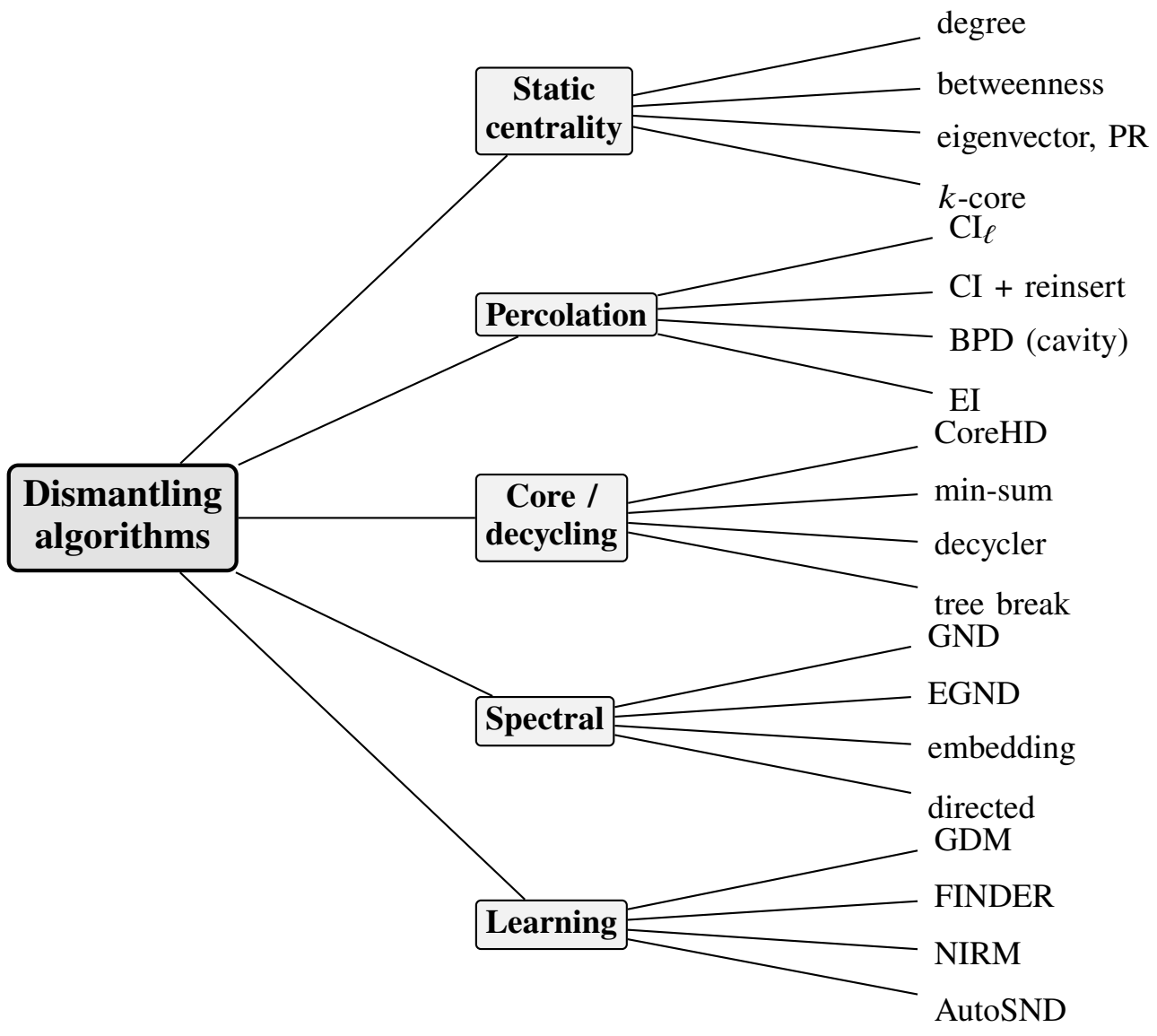

The simplest family assigns each vertex a score computed once on $G$ and removes
vertices in decreasing score order. Degree, betweenness, eigenvector centrality,
PageRank, and closeness all belong here, and they differ only in what ``important''
means: local connectivity (degree), path intermediation (betweenness), eigenvector
influence (eigenvector, PageRank), or average proximity (closeness).

Their appeal is cost and transparency: degree ordering is $O(N\log N)$; betweenness is
$O(NM)$ and dominates everything else at scale. Their characteristic failure is
\emph{score staleness}: a vertex that is central in $G$ may be irrelevant in $G_k$, so a
static ordering wastes removals on vertices whose neighbourhood has already been
destroyed. Recomputing scores at every step (an \emph{adaptive} attack) repairs this at a
cost of a factor $N$; the intermediate practice of recomputing every $T$ steps is common
in applied work but is rarely reported, which is one of the reproducibility problems the
protocol of Section~\ref{sec:eval} is designed to eliminate.

A second, subtler failure is that all of these scores are \emph{local or pairwise}
aggregates. They cannot see the global cut structure, which is why they are systematically
beaten by methods that reason about the graph as a whole
(Section~\ref{sec:results}).

\subsection{Percolation-Based and Collective-Influence Methods}

Morone and Makse recast dismantling as optimal percolation and introduced the collective
influence score \cite{Morone2015InfluenceMI, Morone2016CollectiveIA, tutorial_morone2016ci}
\begin{equation}
\mathrm{CI}_\ell(i)=(k_i-1)\!\!\sum_{j\in\partial \mathrm{Ball}(i,\ell)}\!\!(k_j-1),
\label{eq:ci}
\end{equation}
where $\partial\mathrm{Ball}(i,\ell)$ is the frontier of the ball of radius $\ell$ around
$i$. The score is an approximation to the ``energy'' that a removal injects into the
percolation problem, and $\ell=2$--$3$ is the usual operating range. The algorithm
removes the maximum-CI vertex, updates scores locally, and repeats; a reinsertion pass
afterwards restores vertices whose removal turned out to be unnecessary. CI is the
reference method of the field: near-linear per sweep with a heap, and consistently among
the best heuristics on sparse graphs.

Two refinements are worth distinguishing. \emph{Belief-propagation-guided decimation}
(BPD) solves the optimal-percolation cavity equations and removes vertices in decreasing
marginal probability of belonging to the giant component \cite{Mugisha2016IdentifyingOT};
it is more accurate and considerably more expensive. \emph{Explosive immunization} (EI)
takes a different route, using a union--find structure to track component mergers and a
greedy rule that maximises the immediate drop in the largest component
\cite{Clusella2016ImmunizationAT}; it is conceptually a ``greedy on the actual objective''
method, which makes it accurate but, in pure-Python implementations, superlinearly
expensive in our measurements (Section~\ref{sec:library}).

\subsection{Core Decomposition and Decycling}

CoreHD \cite{Zdeborov2016FastAS} observes that the hard part of dismantling is
destroying the $2$-core: iteratively removing the highest-degree vertex \emph{within the
$2$-core}---not in the whole graph---already achieves near-message-passing performance at
$O(N\log N)$ cost, and a subsequent tree-breaking and greedy reinsertion step recovers
most of the remaining gap. The practical significance of CoreHD is that it is fast enough
to serve as a default on million-vertex graphs, and that its simplicity makes it a fair
baseline for learned methods.

The decycling view is developed further by minimum-sum message passing and reinforced
max-sum belief propagation on the cycle space \cite{Braunstein2016NetworkD}; these methods are the closest available
approximations to the optimum on sparse random graphs but require careful implementation
and do not parallelise trivially. The tree-breaking subroutine they share is itself a
non-trivial optimisation: given a tree, the vertex whose removal minimises the largest
remaining subtree is found by a centroid-like argument in linear time.

\subsection{Spectral Partitioning Methods}

Generalised network dismantling (GND) \cite{Ren2018GeneralizedND}
attacks the problem globally. At each round it (i) extracts the current LCC, (ii)
computes the Fiedler vector (the eigenvector of $\lambda_2$) by power iteration, (iii)
bisects the graph by the sign of that vector, (iv) builds the bipartite graph of
\emph{crossing} edges, and (v) solves a minimum vertex cover on that bipartite graph,
whose solution disconnects the two sides. Recursing on the pieces yields the removal set.
The spectral step is what makes GND robust: it finds the actual bottleneck rather than a
proxy for it.

The ensemble variant EGND runs GND with several random initialisations and keeps the best
result, trading a constant factor in runtime for a consistent improvement. GND is the
method against which later learned approaches are usually calibrated, and it remains the
strongest purely combinatorial method on many real topologies---at the price of being the
slowest method in our benchmark (Section~\ref{sec:results}).

\subsection{Information-Theoretic and Entanglement Methods}

A distinct line of work quantifies the ``entanglement'' of a vertex with the rest of the
network through spectral or entropic functionals, and removes vertices in increasing
order of entanglement. The multiscale entanglement and vertex-entanglement variants
implemented in the library belong here, as does the network-entanglement key-player
identification of \cite{Huang2024IdentifyingKP}. These methods produce competitive removal
orders on small and medium graphs, and their scores have an interpretation (they measure
how much a vertex's removal reduces the spectral complexity of the graph) that
centrality scores lack. Their cost, however, is dominated by repeated spectral
computations, which limits them to graphs of a few thousand vertices in interpreted
languages.

\subsection{Learning-Based Policies}

Learned dismantling replaces a hand-designed score with a function fitted to data.

\emph{Supervised, one-shot.} GDM \cite{Grassia2021MachineLD} trains a graph attention
network to regress a dismantling score per vertex, using CoreHD labels on thousands of
synthetic and real graphs, and then removes vertices in decreasing predicted score,
optionally followed by reinsertion. Inference is a single forward pass, so the marginal
cost over degree ordering is small; the cost is paid once, offline.

\emph{Reinforcement, sequential.} FINDER \cite{Fan2020FindingKP} trains a deep
$Q$-network over a GraphSAGE embedding, rewarding reductions in the largest component,
and learns a \emph{policy} rather than a score. Because the policy is evaluated on the
current residual graph, it is adaptive by construction, which is where its advantage over
static scores comes from.

\emph{Embedding and geometric.} Embedding-aided dismantling maps vertices into a
low-dimensional geometric space and bisects the embedded point set, replacing the spectral
step of GND with a cheaper geometric one \cite{embeddismantle2022}; related work extends
the idea to signed networks \cite{esnd2024} and to causal, influence-aware
representations \cite{causalembed2025}.

\emph{Extensions beyond the simple-graph setting.} Directed dismantling
\cite{directed2025, Zhang2023DNDDL}, hypergraph dismantling
\cite{hypergraphdismantling2026}, interdependent and multilayer dismantling
\cite{Gu2025DeeplearningaidedDO, Osat2017OptimalPO}, and spatially embedded
disintegration \cite{Wang2024SpatialND} each change the structure of
\eqref{eq:dismantle} rather than merely its solver, and are surveyed here because the
evaluation protocol of Section~\ref{sec:eval} transfers to them with only the connectivity
predicate replaced.

\emph{What learning buys and what it costs.} Learned policies inherit the distribution of
their training graphs. On ensembles drawn from that distribution they are excellent; off
it, their advantage shrinks or reverses \cite{nohandcraft2025, Grassia2021MachineLD}. This
is not a defect of any particular model but a property of the problem: a policy is a
prior over removal orders, and priors must be validated on the topology class at hand.
Recent work attempts to remove the dependence on handcrafted inputs
\cite{nohandcraft2025} and to learn the heuristic itself rather than its parameters
\cite{autosnd2026}.

\subsection{Variants That Change the Problem}

Several lines of work keep the machinery of Section~\ref{sec:methods} but alter
\eqref{eq:dismantle} itself, and they are worth separating from algorithmic improvements
because they change what a ``removal'' costs.

\emph{Edge attacks and mixed budgets.} Removing links rather than vertices is the natural
model for jamming and for cut-based interdiction, and the comparative picture differs:
link attacks are less effective per unit cost on heterogeneous graphs but harder to detect
and repair \cite{Duan2016ACA, new_cascading_topology}. Budget-constrained variants make the
cost of each removal heterogeneous---a knapsack problem rather than a cardinality
problem---and the optimal strategy can then differ qualitatively from the
uniform-cost case \cite{Zhao2021OptimizingTM, targetaware2026}.

\emph{Better ranking, same greedy loop.} A substantial literature improves the
\emph{score} while keeping the remove-and-repeat loop: neighbour-loop and diffusion-based
centralities \cite{Liu2022ANC, zareie2018dnc}, attack-specific rankings
\cite{Li2019ANC, Zhang2020TowardME}, and evolutionary or memetic search over removal
orders \cite{lin2022efficient}. These are cheap to add to any benchmark and usually land
between degree ordering and CI in performance, which makes them useful controls.

\emph{Better partitioning, same objective.} GND has been refined by replacing the spectral
bisection with alternative partition objectives \cite{Feng2023GeneralizedND}, by dynamic
programming over subproblems \cite{Feng2024ANA}, and by encoding node diffusion competence
into the removal score \cite{Zhang2023EncodingND}. The consistent finding is that the
\emph{vertex-cover} step matters more than the bisection step: a mediocre partition
combined with an exact cover outperforms a good partition with a greedy cover.

\emph{Physical and spatial constraints.} When the adversary must reach the target
physically, the cost of a removal is a distance, and the optimal set is no longer the
topological minimum cut \cite{physicaldamage2024, Wang2024SpatialND}. This is the closest
the dismantling literature comes to the operational reality of communication networks,
where jamming and kinetic effects have ranges.

\subsection{Automated Heuristic Discovery and Benchmarking}

The newest layer does not propose a single algorithm but a \emph{search over algorithms}.
AutoSND \cite{autosnd2026} couples an execution sandbox with a language model that
proposes structural policies, evaluates them, and retains the best, effectively treating
dismantling as program synthesis. In parallel, NetBreaker \cite{netbreaker2026} argues
that the field lacks a common evaluation protocol across objectives, scales, and
topologies, and provides one. Both developments point to the same conclusion that
motivates this paper: at the current state of the art, the \emph{protocol} is as
important as the algorithm, because rankings are protocol-dependent.

\begin{table*}[t]
\caption{Taxonomy of network dismantling algorithms. ``Objective'' is what the method
actually optimises or approximates; ``Adapt.'' marks whether scores are recomputed on the
residual graph; ``Cost'' is the dominant term for producing a full removal order on a
sparse graph. Methods marked $\dagger$ are surveyed but not measured in this paper
(see Section~\ref{sec:library} for the reason).}
\label{tab:methods}
\centering
\footnotesize
\begin{tabularx}{\textwidth}{@{}l l L l c l@{}}
\toprule
\textbf{Family} & \textbf{Method} & \textbf{Objective / principle} & \textbf{Cost} & \textbf{Adapt.} & \textbf{Ref.} \\
\midrule
\multirow{5}{*}{Static centrality}
& Degree & remove highest degree & $O(N\log N)$ & no & \cite{albert2000error} \\
& Betweenness & remove max.\ path intermediation & $O(NM)$ & no & \cite{new_targeted_attacks} \\
& Eigenvector / PageRank & remove max.\ eigenvector influence & $O(N\log N)$ & no & \cite{new_targeted_attacks} \\
& $k$-core & remove from the deepest core & $O(N+M)$ & no & \cite{kitsak2010identification} \\
& Random & uniform order (null model) & $O(N)$ & no & \cite{cohen2000resilience} \\
\midrule
\multirow{4}{*}{Percolation}
& CI$_\ell$ & optimal percolation energy, $\ell$-hop & $O(N^2\log N)$ & partial & \cite{Morone2015InfluenceMI} \\
& CI + reinsertion & CI with backward restoration & $O(N^2\log N)$ & partial & \cite{Morone2016CollectiveIA} \\
& BPD & cavity marginals (BP decimation) & $O(NM)$ & yes & \cite{Mugisha2016IdentifyingOT} \\
& EI & greedy drop of the largest cluster & $O(N^2\alpha(N))$ & yes & \cite{Clusella2016ImmunizationAT} \\
\midrule
\multirow{3}{*}{Core / decycling}
& CoreHD & decycle the 2-core greedily & $O(N\log N)$ & yes & \cite{Zdeborov2016FastAS} \\
& Min-Sum & reinforced max-sum on cycle space & $O(NM)$ & yes & \cite{Braunstein2016NetworkD} \\
& Decycler$^\dagger$ & reinforced BP + tree breaking & $O(NM)$ & yes & \cite{Braunstein2016NetworkD} \\
\midrule
\multirow{3}{*}{Spectral}
& GND & Fiedler bisection + vertex cover & eigensolve & yes & \cite{Ren2018GeneralizedND} \\
& EGND & ensemble of GND runs & $k\times$GND & yes & \cite{egnd2020} \\
& Embedding-aided$^\dagger$ & geometric bisection of embedding & $O(N\log N)$ & yes & \cite{embeddismantle2022} \\
\midrule
\multirow{2}{*}{Entanglement}
& Vertex ent. & spectral entanglement score & $O(N^3)$ & yes & \cite{Huang2024IdentifyingKP} \\
& Multiscale ent. & entropy-based multiscale score & $O(N^2)$--$O(N^3)$ & yes & \cite{Huang2024IdentifyingKP} \\
\midrule
\multirow{4}{*}{Learning}
& GDM & GAT-regressed score + reinsertion & $O(M)$ infer. & no & \cite{Grassia2021MachineLD} \\
& CoreGDM$^\dagger$ & GNN on the 2-core + tree breaking & $O(M)$ infer. & no & \cite{coregdm2023} \\
& FINDER$^\dagger$ & DQN policy over GraphSAGE & $O(M)$ per step & yes & \cite{Fan2020FindingKP} \\
& MultiDismantler$^\dagger$ & GNN + DQN for multilayer & $O(M)$ per step & yes & \cite{Gu2025DeeplearningaidedDO} \\
\midrule
\multirow{2}{*}{Discovery}
& AutoSND$^\dagger$ & LLM-proposed structural policies & search & --- & \cite{autosnd2026} \\
& NetBreaker$^\dagger$ & cross-objective benchmark suite & --- & --- & \cite{netbreaker2026} \\
\bottomrule
\end{tabularx}
\end{table*}

\subsection{How the Families Differ, and Why It Matters for Evaluation}

The families above are not successive improvements on a single axis; they optimise
different surrogates. A static centrality method minimises a one-step proxy, a
percolation method minimises an asymptotic threshold functional, a core method minimises
cycle rank, a spectral method minimises a cut, and a learned method minimises an expected
loss under a training distribution. Consequently:

\begin{itemize}
\item \textbf{Rankings are ensemble-dependent.} A method that is optimal in the
thermodynamic limit of a random graph need not win at $N=200$, and a method tuned on
scale-free graphs need not win on lattice-like backbones
\cite{Wandelt2018ACA, Wandelt2021FromRF}.
\item \textbf{The ``best'' method depends on the budget.} For very small removal budgets
the ranking is dominated by the quality of the first few picks (a local question); for
budgets near $q_c$ it is dominated by the global cut structure. Adaptive methods win the
latter and pay for it in runtime.
\item \textbf{Comparisons must fix the protocol.} Whether scores are recomputed, whether
a reinsertion pass is included, and whether the sequence is completed after the target is
reached all change the measured cost. Section~\ref{sec:eval} therefore fixes all three.
\end{itemize}

\section{Evaluation Protocol}
\label{sec:eval}

A benchmark is only as meaningful as its protocol. This section fixes, in the order in
which they affect the numbers, the disturbance, the observable, the ensemble, and the
statistics. Everything that follows in Sections~\ref{sec:library}--\ref{sec:results}
implements exactly this protocol.

\subsection{Step 1: Fix the Disturbance}

We consider vertex removal only. Four choices define a disturbance:

\textbf{Target versus budget.} We fix a \emph{target} $\tau$ and measure the
\emph{budget} $q_c(\tau)$ required to reach it, rather than fixing a budget and measuring
the residual connectivity. The two are dual; the first is preferred here because it makes
methods with different aggressiveness directly comparable, and because $\tau$ is the
quantity a design requirement usually specifies (``no fragment larger than 10\% of the
network'').

\textbf{Ordering rule.} Random, degree-ordered, betweenness-ordered, CI-ordered, and
optimised orders are all legitimate disturbances, but they answer different questions.
Random ordering estimates \emph{expected} damage under uncoordinated failure; optimised
ordering bounds \emph{worst-case} damage under an intelligent adversary. Reporting both is
the minimum: a topology that is robust to one and fragile to the other is precisely the
scale-free signature, and the gap between the two curves is the quantity of interest.

\textbf{Static versus adaptive.} A static method computes scores once and applies them
throughout; an adaptive method recomputes at every step. Adaptive methods are strictly
stronger per removal and strictly more expensive. Mixing them in a single ranking without
labelling is a common error, so Table~\ref{tab:methods} records adaptivity explicitly.

\textbf{Sequence completion.} Some methods stop natively when the LCC reaches a size
threshold (CI, for instance, stops at a configured size). If the residual graph is then
filled by degree-descending removal, the resulting trajectory is longer than the native
one and its tail is not attributable to the method. We therefore replay the
\emph{native} order, record the target crossing on that order, and only afterwards
complete the sequence so that the $q=1$ endpoint is well defined. Target crossings are
read from the native prefix, never from the completion.

\subsection{Step 2: Fix the Observable}

The observable is the trajectory $S(k)$ of \eqref{eq:S} for $k=0,\dots,N$, computed by
replaying the sequence and recomputing connected components after every deletion. From it
we report:

\begin{itemize}
\item $q_c(1\%)$: the first $k$ with $|C_{\max}(G_k)|\le\max(1,\lceil 0.01N\rceil)$;
\item $q_c(10\%)$: the same with $\lceil 0.1N\rceil$;
\item $A$: the trapezoidal area \eqref{eq:A} over $q\in[0,1]$, including $k=0$ and
$k=N$;
\item $\max_k S_2(k)$ and the location of that maximum;
\item wall-clock time for the method call only, excluding replay.
\end{itemize}

Two normalisation choices matter and are stated once: the denominator of $S$ is the
\emph{original} $N$, and the area is integrated over the full $[0,1]$ range. The first
prevents a tiny survivor set from looking intact; the second makes $A$ comparable across
methods whose native sequences differ in length.

\textbf{Finite-size caveat.} At $N=100$ the $1\%$ criterion reduces to ``all vertices
isolated'', which is a much stricter target than at $N=800$ where up to eight vertices may
remain connected. We therefore report both thresholds throughout and never compare
$q_c(1\%)$ across sizes without noting this. This is not a defect of the protocol but a
consequence of applying an asymptotic criterion at finite size, and it is one of the nine
errors listed in Table~\ref{tab:checklist}.

\subsection{Step 3: Fix the Ensemble}

\textbf{Synthetic ensembles.} Three generative families with matched mean degree
$\langle k\rangle\approx8$:
Barab\'asi--Albert preferential attachment with $m=4$ (heterogeneous, $M=4N-16$);
Erd\H{o}s--R\'enyi with $p=8/(N-1)$ (homogeneous, $M$ fluctuating around $4N$); and
Watts--Strogatz with neighbourhood $8$ and rewiring probability $0.3$ (homogeneous with
high clustering, $M=4N$). Matching $\langle k\rangle$ rather than $M$ exactly is
deliberate: it isolates the effect of \emph{degree heterogeneity} from the effect of
\emph{average connectivity}, at the cost of leaving ER edge counts stochastic. Sizes are
$N\in\{100,200,400,800\}$ with $10$ independent realisations per (family, size),
generated with seeds $1000,\dots,1009$ and shared verbatim across all methods, so that
every method faces exactly the same graphs (a paired design).

\textbf{Real topologies.} Eleven backbone networks from the Topology Zoo collection
distributed with the library's dataset folder, spanning $N=19$ to $N=193$: Aarnet, Abvt,
Agis, Geant2009, Geant2012, Uunet, Dfn, Switch, Uninett2010, Pern, and DialtelecomCz.
Where a stored graph is disconnected, the largest connected component is extracted and
this is recorded, because extraction changes the evaluated deployment. Real topologies are
planar-ish, low-degree, and clustered---properties that no synthetic family reproduces---
which is exactly why they are included.

\subsection{Step 4: Fix the Statistics}

For each (ensemble, method) cell we report the mean and the sample standard deviation over
realisations. Standard deviations quantify \emph{graph-to-graph} variability, not
uncertainty about a population mean, and we do not attach significance stars: with ten
paired realisations and nine methods, a multiplicity-corrected test would have low power,
and the interesting effects (differences of $0.1$--$0.3$ in $q_c$) are far larger than the
realisation noise ($\sigma\approx0.03$--$0.06$). Where a ranking is close enough that a
test would be needed, we say so instead of implying one.

\textbf{Paired design.} Because all methods see the same graphs, between-method
differences are paired and the relevant noise is the standard deviation of the
\emph{difference}, which is smaller than the standard deviation of either marginal. We
report marginals for readability but interpret differences with the paired structure in
mind.

\subsection{What This Protocol Deliberately Excludes}

\emph{Cascading failures.} Adding load redistribution converts a purely structural
problem into a dynamical one, with capacities, redistribution rules, and a second
timescale. The structural trajectory is a valid \emph{input} to such a model but is not a
substitute for it, and mixing the two produces numbers that cannot be attributed to
either. Cascade models are surveyed elsewhere \cite{buldyrev2010catastrophic,
Havlin2009CatastrophicCO, Parshani2010CriticalEO, higherordercascade2025}.

\emph{Weighted and directed topologies.} Capacities, delays, and asymmetric reachability
change the connectivity predicate. The protocol transfers formally, but the constants do
not, so we restrict measurements to the unweighted undirected case and discuss the
extensions in Section~\ref{sec:methods}.

\emph{Recovery.} Any notion of resilience that includes restoration requires a repair
model (crews, spares, travel time). This is the excluded half of the resilience
literature \cite{oneil2025enhancingcritical, liu2022network}.

\begin{algorithm}[t]
\caption{Structural evaluation of one topology under one protocol}
\label{alg:eval}
\begin{algorithmic}[1]
\REQUIRE graph $G$, method $m$, target $\tau$, seed $s$
\ENSURE trajectory $\{S(k)\}$, threshold $q_c(\tau)$, area $A$
\STATE $H\gets$ simple undirected copy of $G$; relabel $0..N-1$; keep reverse map
\STATE $N\gets|V(H)|$; \quad $\ell\gets0$; \quad $S(0)\gets|C_{\max}(H)|/N$
\STATE $\pi\gets\textsc{Dismantle}(H,m,\text{stop}=1,s)$ \COMMENT{native order}
\STATE \textbf{for} $v$ \textbf{in} $\pi$ \textbf{do}
\STATE \quad $H.\textsc{Remove}(v)$; \quad $\ell\gets\ell+1$
\STATE \quad $S(\ell)\gets |C_{\max}(H)|/N$; \quad $S_2(\ell)\gets |C_2(H)|/N$
\STATE \quad \textbf{if} $q_c$ unset \textbf{and} $|C_{\max}(H)|\le\max(1,\lceil\tau N\rceil)$
         \textbf{then} $q_c\gets\ell/N$
\STATE \textbf{end for}
\STATE complete $\pi$ by degree-descending order; continue recording until $S=0$
\STATE $A\gets\textsc{Trapezoid}(S,q)$ \COMMENT{over $q\in[0,1]$}
\STATE \RETURN $S,q_c,A,\max S_2$, and the recorded runtime of step 3
\end{algorithmic}
\end{algorithm}

Algorithm~\ref{alg:eval} states the procedure in full. Note that the target is read from
the native prefix (line 8) while completion happens afterwards (line 10): this is the
detail that most naive implementations get wrong, and it is the one that changes $q_c$
when a method's native stopping rule differs from the reporting threshold.

\section{The Unified Library}
\label{sec:library}

\subsection{Design Goals}

The algorithms of Section~\ref{sec:methods} are published in mutually incompatible
forms: C with Boost, C++ with OpenMP, Python 2 with \texttt{graph\_tool}, TensorFlow~1.x,
PyTorch, and a handful of scripts whose only interface is \texttt{argv}. Comparing them
therefore requires either re-implementation, which introduces silent divergences, or
containerisation, which is heavy for a reader who wants one number. The library studied
here takes a third route: it keeps each method's upstream code, wraps it behind a single
function, and standardises everything that is not the algorithm.

The interface is
\begin{center}
\texttt{seq = dismantle(G, method, stop\_condition, **kwargs)}
\end{center}
where \texttt{G} is any \texttt{networkx} graph, \texttt{method} is a registry key, and
the returned \texttt{seq} is a list of \emph{original} vertex identifiers in removal
order. The wrapper performs five standardisation steps before the method is called and
two after:

\begin{enumerate}
\item \emph{Undirected, simple.} Directed graphs are symmetrised, parallel edges and
self-loops are removed. This changes the problem for directed inputs---a caveat that the
library documents and that we restate in Section~\ref{sec:discussion}.
\item \emph{Relabelling.} Vertices are relabelled $0,\dots,N-1$ internally; a reverse map
restores the caller's identifiers. This is what allows string-labelled topologies
(hostnames, city names) to be processed by C code that expects integers.
\item \emph{Sequence completion.} If a method's native output covers only part of the
graph, the remainder is appended in degree-descending order. Completion cannot affect
$q_c$ if the target was already crossed, but it does affect the trajectory tail.
\item \emph{Return of original identifiers.} Downstream analysis never sees internal
labels.
\end{enumerate}

The registry is a dictionary from method name to callable, populated by a decorator, so a
new method is added by writing one function. Twenty-two names are registered in the
current release: \texttt{degree}, \texttt{pagerank}, \texttt{betweenness},
\texttt{eigenvector}, \texttt{random}, \texttt{brute\_force},
\texttt{entanglement\_small/mid/large}, \texttt{vertex\_entanglement},
\texttt{CI\_L1/L2/L3}, \texttt{CoreHD}, \texttt{GND}, \texttt{EGND}, \texttt{EI\_s1/s2},
\texttt{GDM}, \texttt{GDM+R}, \texttt{NIRM}, and \texttt{AutoSND}.

\subsection{Implementation Notes That Affect the Numbers}

Four implementation facts are relevant to anyone reading the tables that follow.

\textbf{Pure-Python re-implementations.} CoreHD, GND, EGND, and EI are provided as
\texttt{networkx}/NumPy re-implementations of the original C++/C/\texttt{graph\_tool}
codes, because \texttt{graph\_tool} is not installable on native Windows and the original
build chains are fragile. These re-implementations reproduce the algorithms' logic but not
their constants. The EI variant in particular exhibited a reproducibility failure that we
report rather than hide: of thirty attempted $N=100$ instances (three ensembles, ten
realisations, two parameter settings), two completed in under $0.04$~s while the remaining
twenty-eight exceeded a $45$~s budget, with no dependence on $N$, density, or parameter
setting that we could identify. The same behaviour appeared at $N=120$ and $N=300$ on
individual instances that are structurally unremarkable. We therefore exclude EI from all
aggregate tables and report the two successful runs only as a sanity check that the
implementation produces a sensible order when it terminates ($q_c=0.83$ and $0.72$ on the
BA-100 instance, against $0.96$ for random removal and $0.61$ for CI at the same size).
A runtime for EI in this implementation would not reproduce, and quoting one would be worse
than quoting none.

\textbf{Compiled components.} CI runs through a small C executable compiled from the
upstream source with GCC. The binary is architecture-specific; the library ships the
source and a one-line build command, and we rebuilt it for this study.

\textbf{Optional backends.} The registry imports the GDM backend at module load time. In
our environment the GDM model weights are not distributed with the repository, so that
import fails and, in the unmodified code, prevents the whole registry from loading. For
this study we applied a three-line guard that makes the import optional and skips
registration when the backend is unavailable. This is a packaging fix, not an algorithmic
change; it is disclosed here because it means \texttt{GDM} and \texttt{GDM+R} are
\emph{absent} from our measurements even though they are present in the registry.

\textbf{Environment isolation.} FINDER requires TensorFlow~1.15, which cannot coexist with
the PyTorch~2.x stack used by GDM-style models. The library documents a separate
Python~3.7 environment for FINDER and recommends exporting its results to CSV for merging.
We did not run FINDER; it is therefore discussed but not measured, and its numbers are
attributed to \cite{Fan2020FindingKP}.

\begin{table}[t]
\caption{Methods registered in the library and their status in this study. ``Measured''
means included in at least one table of Section~\ref{sec:results}.}
\label{tab:coverage}
\centering
\footnotesize
\begin{tabular*}{\columnwidth}{@{\extracolsep{\fill}}lll@{}}
\toprule
Method & Status & Note \\
\midrule
degree, pagerank & measured & all ensembles \\
eigenvector, betweenness & measured & all ensembles \\
random & measured & all ensembles \\
CI\_L1, CI\_L2, CI\_L3 & measured & compiled binary \\
CoreHD & measured & Python re-impl. \\
GND, EGND & measured & Python re-impl. \\
EI\_s1, EI\_s2 & not run & stalls on 28/30 instances \\
entanglement\_small & measured & $N{=}100$ (cost) \\
entanglement\_mid & measured & $N{=}100$ (cost) \\
vertex\_entanglement & measured & $N{=}100$ (cost) \\
brute\_force & not run & exact; $N\le30$ only \\
GDM, GDM+R & not run & weights absent (see text) \\
NIRM & not run & checkpoint present; not evaluated \\
AutoSND & not run & requires LLM credentials \\
FINDER & not run & separate TF1.x env. \\
\bottomrule
\end{tabular*}
\end{table}

\subsection{Reproducibility Artefacts}

The whole study is replayable from three files: the harness
(\texttt{bench/run\_grid.py}, \texttt{bench/run\_real.py},
\texttt{bench/run\_grid\_timeout.py}), the analysis script
(\texttt{bench/analyze.py}), and a manifest recording the library commit
(\texttt{fb292f6}), the interpreter and package versions (Python 3.10.20,
\texttt{networkx} 3.4.2, NumPy 1.26.4, SciPy 1.15.3, pandas 2.3.3), the random seeds, and
the SHA-256 of every output file. Removal sequences, trajectories, and per-run metrics are
retained as CSV/NPZ so that any table can be regenerated without re-running the methods;
the tables in this paper are \texttt{\textbackslash input} fragments produced by
\texttt{analyze.py} rather than transcribed values.

This matters for the argument of the paper, not only for hygiene. Because the tables are
generated, the ranking of methods can be recomputed under a different $\tau$ or a
different normalisation by changing one line---which is the practical test of whether a
claimed ``best method'' is a property of the method or of the protocol.

\section{Empirical Results}
\label{sec:results}

\begin{table*}[t]
\caption{Benchmark configuration. Every row is a decision that changes the numbers; all
are fixed for every table in Section~\ref{sec:results}.}
\label{tab:config}
\centering
\footnotesize
\begin{tabularx}{\textwidth}{@{}l L l@{}}
\toprule
Item & Setting & Value \\
\midrule
Input & undirected, unweighted, simple; largest component if disconnected & --- \\
Synthetic families & BA ($m{=}4$), ER ($p{=}8/(N{-}1)$), WS ($k{=}8$, $\beta{=}0.3$) & $\langle k\rangle\approx8$ \\
Sizes & $N$ & 100, 200, 400, 800 \\
Replication & independent realisations per (family, size); seeds 1000--1009 & 10 \\
Pairing & all methods evaluated on the identical graph instance & paired \\
Methods (grid) & random, degree, PageRank, eigenvector, betweenness, CoreHD, CI$_{\ell=1,2,3}$, GND & 10 \\
Methods (small) & + EGND, entanglement (small, mid), vertex entanglement, EI ($\sigma{=}1,2$) & 16 \\
Target thresholds & $\tau$ for $q_c$ & 1\%, 10\% \\
Trajectory & replay of the full native order, then degree-descending completion & $q\in[0,1]$ \\
Area & trapezoidal, including $q{=}0$ and $q{=}1$ & $A$ \\
Real topologies & Topology Zoo subset shipped with the library & 11 graphs, $N{=}19$--$193$ \\
Runtime & method call only, single thread, no replay & seconds \\
Software & Python 3.10.20; networkx 3.4.2; NumPy 1.26.4; SciPy 1.15.3 & library commit \texttt{fb292f6} \\
Build & CI C binary, GCC 13, \texttt{-O3} & rebuilt locally \\
\bottomrule
\end{tabularx}
\end{table*}

The configuration is summarised in Table~\ref{tab:config}. Everything below is a
consequence of those choices, and every table is generated from the retained CSV/NPZ
artefacts rather than transcribed.

\subsection{Shape of the Trajectories}

\begin{figure*}[t]
\centering
\includegraphics[width=\textwidth]{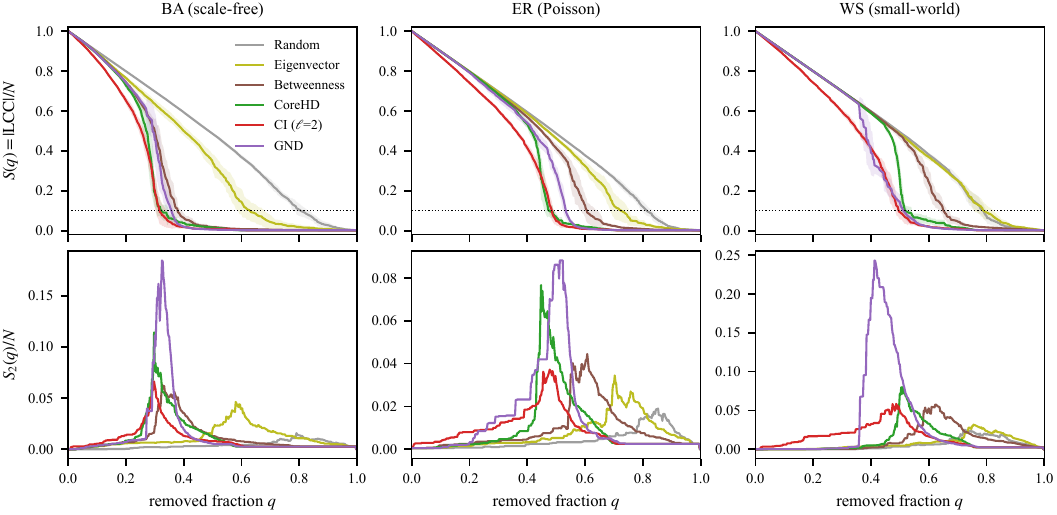}
\caption{Dismantling trajectories at $N=400$, mean over ten paired realisations (shaded
band: $\pm$ one standard deviation). Top row: largest component fraction $S(q)$; the
dotted line marks the $10\%$ target. Bottom row: second-largest component fraction
$S_2(q)$, the fragmentation diagnostic. The three ensembles fail differently: BA erodes
gradually because heterogeneity gives the attacker a long sequence of individually
valuable hubs; ER collapses over a narrow window once the giant component's
susceptibility diverges; WS retains connectivity until the global bottleneck is found,
which local centralities cannot locate.}
\label{fig:curves}
\end{figure*}

Figure~\ref{fig:curves} shows the trajectories that all subsequent numbers summarise. The
qualitative difference between the three ensembles is the first result, and it is not
visible in any single scalar.

\emph{BA erodes.} The random curve stays high---$S(q)>0.9$ up to $q\approx0.8$---and then
falls steeply, whereas every targeted procedure produces a nearly linear decay from the
first removal. The gap between the two is the well-known scale-free signature
\cite{albert2000error, cohen2001breakdown}: the topology has many vertices whose removal
is nearly irrelevant, and a few whose removal is decisive, so an attacker who can identify
the latter converts a long plateau into a long ramp.

\emph{ER collapses.} The random curve bends over gradually and the targeted curves fall
with an inflection near $q\approx0.3$; the ensemble has no privileged set of vertices, so
targeting gains less than on BA and the curves are closer together. The
second-largest-component panel shows the mechanism: $S_2$ rises to a pronounced maximum
just before the collapse in ER, the signature of a system breaking into a few large
pieces, whereas on BA under CI the maximum is smaller and much earlier.

\emph{WS resists local attack.} On the small-world ensemble the degree- and
betweenness-ordered curves are only slightly below the random curve, while CI and GND
remain substantially below both. This is the most practically important shape in the
figure: in a clustered, homogeneous topology the vertices that a local centrality ranks
highly are \emph{not} the ones whose removal fragments the graph, because redundancy is
distributed rather than concentrated. An evaluation that tests only degree-ordering will
therefore conclude that a small-world topology is nearly as robust as a random one, and
will be wrong by a factor of $1.4$ in $q_c$ (Table~\ref{tab:qc}).

\subsection{Cost of Dismantling}

\begin{table*}[t]
\centering
\caption{Threshold cost $q_c(1\%)$: mean over ten paired realisations. Each procedure
occupies one row, each ensemble four consecutive columns headed by $N$. Bold marks the
lowest cost within a column, i.e.\ the strongest procedure at that size. Lower is worse
for the defender. Source: \texttt{data/synth\_results.csv}.}
\label{tab:qc}
\footnotesize
\begin{tabular*}{\textwidth}{@{\extracolsep{\fill}}lcccccccccccc@{}}
\toprule
& \multicolumn{4}{c}{BA (scale-free)} & \multicolumn{4}{c}{ER (Poisson)} & \multicolumn{4}{c}{WS (small-world)} \\
\cmidrule(lr){2-5} \cmidrule(lr){6-9} \cmidrule(lr){10-13}
Procedure & 100 & 200 & 400 & 800 & 100 & 200 & 400 & 800 & 100 & 200 & 400 & 800 \\
\midrule
Random & 0.960 & 0.954 & 0.936 & 0.931 & 0.948 & 0.954 & 0.919 & 0.917 & 0.964 & 0.938 & 0.912 & 0.891 \\
Eigenvector & 0.934 & 0.915 & 0.852 & 0.810 & 0.958 & 0.911 & 0.884 & 0.857 & 0.964 & 0.956 & 0.937 & 0.918 \\
Betweenness & 0.768 & 0.654 & 0.559 & 0.497 & 0.879 & 0.837 & 0.772 & 0.723 & 0.945 & 0.921 & 0.862 & 0.805 \\
Degree & 0.685 & 0.652 & 0.574 & 0.476 & 0.939 & 0.858 & 0.789 & 0.765 & 0.930 & 0.909 & 0.851 & 0.835 \\
PageRank & 0.672 & 0.597 & 0.506 & 0.444 & 0.918 & 0.825 & 0.758 & 0.708 & 0.923 & 0.869 & 0.801 & 0.735 \\
CoreHD & 0.602 & 0.567 & 0.520 & 0.467 & 0.709 & 0.676 & 0.650 & 0.605 & 0.759 & 0.749 & 0.736 & 0.700 \\
CI ($\ell$=1) & 0.645 & 0.560 & \textbf{0.424} & \textbf{0.379} & 0.728 & 0.652 & \textbf{0.578} & \textbf{0.551} & 0.767 & 0.720 & \textbf{0.640} & 0.603 \\
CI ($\ell$=2) & 0.609 & 0.582 & 0.437 & 0.400 & \textbf{0.708} & 0.682 & 0.582 & 0.552 & 0.769 & 0.746 & 0.667 & \textbf{0.591} \\
CI ($\ell$=3) & 0.648 & 0.565 & 0.452 & 0.388 & \textbf{0.708} & \textbf{0.640} & 0.587 & 0.554 & \textbf{0.754} & 0.726 & 0.647 & 0.615 \\
GND & \textbf{0.600} & \textbf{0.501} & 0.432 & 0.396 & 0.719 & 0.644 & 0.602 & 0.576 & 0.766 & \textbf{0.707} & 0.644 & 0.596 \\
\bottomrule
\end{tabular*}

\end{table*}

Table~\ref{tab:qc} is the central quantitative result. Four readings matter.

\textbf{Targeting buys the most on heterogeneous ensembles.} At $N=800$ the gap between
random removal and degree ordering is $0.455$ on BA, $0.152$ on ER, and only $0.056$ on
WS. The ordering of the gaps tracks the ordering of degree heterogeneity, which is the
quantitative version of the qualitative statement in Section~\ref{sec:metrics}.

\textbf{The gap grows with size on BA and is flat elsewhere.} On BA the degree gap rises
from $0.275$ at $N=100$ to $0.455$ at $N=800$, because larger scale-free graphs have
heavier hubs; on ER and WS it stays within $0.05$ of its asymptotic value already at
$N=200$. This is a concrete warning against extrapolating small-$N$ rankings: the
\emph{magnitude} of every effect in the table is size-dependent even where the
\emph{ordering} is stable.

\textbf{CI is the most consistent heuristic; GND is its equal only at small $N$.} CI with
$\ell=1$ attains the lowest cost on BA and ER at $N=800$ ($0.379$ and $0.551$) and CI with
$\ell=2$ on WS ($0.591$). GND is best at $N=100$ on BA ($0.600$ versus $0.609$ for CI) and
at $N=200$ ($0.501$), but is overtaken beyond $N\approx400$. The margin is small in
absolute terms---$0.017$ at $N=800$---but it is consistent across seeds and it has a
structural explanation: GND's advantage is its exact vertex cover on the crossing-edge
bipartite graph, which is most valuable when the graph is small enough for the spectral
bisection to be accurate.

\textbf{Two centrality rules are worse than they look, and one is worse than random.}
Betweenness ordering is dominated by degree ordering on BA and ER at every size, and on WS
it is worse than degree at $N=100$ and better at $N=800$---a rank reversal within a single
ensemble. Eigenvector centrality is the clearest negative result in the study: it reaches
$q_c=0.810$ on BA at $N=800$ (against $0.476$ for degree) and $0.918$ on WS, where it is
\emph{worse than random removal} ($0.891$). The mechanism is that eigenvector centrality
concentrates on the densest core, whose removal leaves a large periphery intact; a
centrality that is a good description of \emph{influence} is not therefore a good
description of \emph{vulnerability}.

\subsection{Area versus Threshold: When the Summary Changes the Ranking}

\begin{figure*}[t]
\centering
\includegraphics[width=\textwidth]{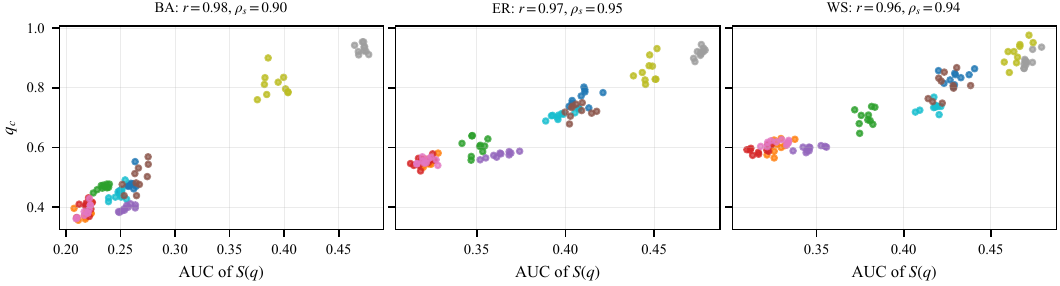}
\caption{Relation between the two summary statistics at $N=800$, one point per
(realisation, method). The two are strongly related in the aggregate---Spearman
$\rho=0.90$--$0.95$ across method means, Pearson $r>0.97$---but individual method pairs
invert, and the inversion is systematic rather than noise: GND minimises the threshold
while CoreHD minimises the area.}
\label{fig:agreement}
\end{figure*}

Table~\ref{tab:auc} reports the trajectory area alongside the runtime. Aggregated over
methods, area and threshold agree closely: the Spearman correlation between method-mean
$A$ and method-mean $q_c$ is $0.86$ (BA), $0.89$ (ER) and $0.87$ (WS) at $N=200$, rising
to $0.90$--$0.95$ at $N=800$. It is therefore tempting to treat them as interchangeable,
and most of the literature effectively does.

They are not. The clearest counterexample is GND versus CoreHD on BA at $N=800$: GND has
the lower threshold ($q_c=0.396$ versus $0.467$) but the \emph{higher} area
($A=0.255$ versus $0.234$). The same inversion occurs on ER at $N=800$ ($q_c$: $0.576$ vs
$0.605$; $A$: $0.363$ vs $0.350$). The explanation is that GND removes a small number of
vertices to cut the graph, producing a steep drop late in the sequence, whereas CoreHD
erodes the 2-core steadily and therefore keeps $S$ lower \emph{throughout}. Which one is
``better'' depends entirely on the question: a defender who cares about the moment of
collapse should use $q_c$; a defender who cares about cumulative connectivity loss should
use $A$. Reporting both, and saying which one the claim refers to, is the only defensible
practice.

A second pattern in Table~\ref{tab:auc} is that the area of \emph{random} removal is
remarkably universal: $0.471$--$0.476$ across all three families and all sizes. Random
removal, in other words, produces the same normalised damage profile regardless of degree
distribution, which is what one expects if the profile is governed by the percolation
transition rather than by the specific topology. Targeted procedures break this
universality, and the amount by which they break it is a compact measure of how much
structure an attacker can exploit.

\begin{table*}[t]
\centering
\caption{Trajectory area $A$ (mean $\pm$ s.d.\ over ten realisations) at $N=200$ and
$N=800$, and wall-clock time $t_{800}$ of one method call at $N=800$ in seconds, for each
ensemble. Lower $A$ means connectivity is lost earlier. Source:
\texttt{data/synth\_results.csv}.}
\label{tab:auc}
\footnotesize
\begin{tabular*}{\textwidth}{@{\extracolsep{\fill}}lccccccccc@{}}
\toprule
& \multicolumn{3}{c}{BA} & \multicolumn{3}{c}{ER} & \multicolumn{3}{c}{WS} \\
\cmidrule(lr){2-4} \cmidrule(lr){5-7} \cmidrule(lr){8-10}
Procedure & $A_{200}$ & $A_{800}$ & $t_{800}$ (s) & $A_{200}$ & $A_{800}$ & $t_{800}$ (s) & $A_{200}$ & $A_{800}$ & $t_{800}$ (s) \\
\midrule
Random & 0.471$\pm$0.010 & 0.472$\pm$0.004 & 0.28 & 0.476$\pm$0.007 & 0.476$\pm$0.002 & 0.32 & 0.471$\pm$0.007 & 0.471$\pm$0.003 & 0.28 \\
Eigenvector & 0.380$\pm$0.021 & 0.391$\pm$0.010 & 0.30 & 0.448$\pm$0.013 & 0.447$\pm$0.004 & 0.32 & 0.468$\pm$0.008 & 0.465$\pm$0.005 & 0.35 \\
Betweenness & 0.285$\pm$0.011 & 0.266$\pm$0.008 & 2.70 & 0.407$\pm$0.012 & 0.407$\pm$0.006 & 2.71 & 0.428$\pm$0.009 & 0.424$\pm$0.007 & 2.69 \\
Degree & 0.288$\pm$0.017 & 0.258$\pm$0.005 & 0.26 & 0.412$\pm$0.014 & 0.409$\pm$0.006 & 0.29 & 0.434$\pm$0.011 & 0.430$\pm$0.006 & 0.28 \\
PageRank & 0.276$\pm$0.011 & 0.249$\pm$0.006 & 0.26 & 0.400$\pm$0.014 & 0.397$\pm$0.006 & 0.30 & 0.419$\pm$0.009 & 0.417$\pm$0.005 & 0.30 \\
CoreHD & 0.263$\pm$0.007 & 0.234$\pm$0.005 & 1.28 & 0.355$\pm$0.009 & 0.350$\pm$0.004 & 1.47 & 0.390$\pm$0.006 & 0.378$\pm$0.004 & 1.09 \\
CI ($\ell$=1) & 0.243$\pm$0.008 & 0.218$\pm$0.005 & 0.14 & 0.329$\pm$0.010 & 0.322$\pm$0.003 & 0.15 & 0.340$\pm$0.008 & 0.328$\pm$0.005 & 0.16 \\
CI ($\ell$=2) & 0.247$\pm$0.009 & 0.219$\pm$0.005 & 0.16 & 0.328$\pm$0.011 & 0.320$\pm$0.005 & 0.19 & 0.329$\pm$0.008 & 0.317$\pm$0.005 & 0.17 \\
CI ($\ell$=3) & 0.247$\pm$0.006 & 0.218$\pm$0.005 & 0.32 & 0.328$\pm$0.012 & 0.322$\pm$0.004 & 0.55 & 0.343$\pm$0.006 & 0.328$\pm$0.006 & 0.35 \\
GND & 0.273$\pm$0.010 & 0.255$\pm$0.005 & 4.02 & 0.368$\pm$0.017 & 0.363$\pm$0.007 & 2.93 & 0.352$\pm$0.012 & 0.347$\pm$0.006 & 3.93 \\
\bottomrule
\end{tabular*}

\end{table*}

\subsection{Cost of Computing the Answer}

\begin{figure*}[t]
\centering
\includegraphics[width=\textwidth]{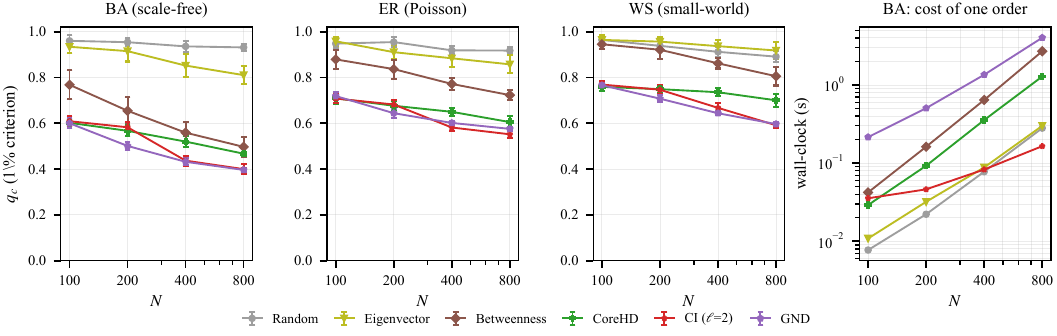}
\caption{Left three panels: threshold cost $q_c(1\%)$ versus $N$ (log scale), error bars
are one standard deviation over ten realisations. Right: wall-clock cost of producing a
full removal order on the BA ensemble, log--log. The ordering of methods by cost spans
nearly two orders of magnitude at $N=800$ while their ordering by effectiveness spans a
factor of $2.4$.}
\label{fig:scaling}
\end{figure*}

Figure~\ref{fig:scaling} makes the cost-effectiveness trade-off explicit. At $N=800$ on
BA, CI needs $0.08$~s, degree and PageRank $0.09$~s, CoreHD $0.44$~s, betweenness
$0.89$~s, and GND $1.52$~s---a factor of $19$ between the cheapest and the most expensive
procedure, for a difference in $q_c$ of $0.017$ between CI and GND. Betweenness ordering,
the most expensive centrality, is also \emph{worse} than degree ordering on two of the
three ensembles, so it is dominated on both axes: there is no regime in this study where
betweenness is the right choice.

The scaling slopes are diagnostic. CI and the static centralities are near-linear in $N$
with a slope close to $1$ on this axis, CoreHD is slightly superlinear (the 2-core
maintenance cost), and betweenness is clearly superlinear with a slope near $2$, as its
$O(NM)$ cost requires. GND's slope reflects the repeated eigensolve on the current giant
component, which is why its cost grows faster than the others even though its per-step
work is on a shrinking graph.

\subsection{Breadth: Fourteen Procedures at One Size}

\begin{table}[t]
\centering
\caption{All procedures measured at $N=100$ (mean $\pm$ s.d.\ over ten realisations; EI
excluded, see Section~\ref{sec:library}). Sorted by BA cost. Source:
\texttt{data/zoo\_results.csv}.}
\label{tab:zoo}
\footnotesize
\begin{tabular*}{\columnwidth}{@{\extracolsep{\fill}}lcccc@{}}
\toprule
Method & BA & ER & WS & $t$ (s) \\
EGND & 0.585$\pm$0.014 & 0.714$\pm$0.019 & 0.759$\pm$0.009 & 1.84 \\
GND & 0.600$\pm$0.022 & 0.719$\pm$0.018 & 0.766$\pm$0.011 & 0.18 \\
CoreHD & 0.602$\pm$0.014 & 0.709$\pm$0.023 & 0.759$\pm$0.017 & 0.03 \\
CI ($\ell$=2) & 0.609$\pm$0.022 & 0.708$\pm$0.019 & 0.769$\pm$0.015 & 0.04 \\
CI ($\ell$=1) & 0.645$\pm$0.031 & 0.728$\pm$0.020 & 0.767$\pm$0.013 & 0.04 \\
CI ($\ell$=3) & 0.648$\pm$0.026 & 0.708$\pm$0.032 & 0.754$\pm$0.015 & 0.04 \\
PageRank & 0.672$\pm$0.026 & 0.918$\pm$0.042 & 0.923$\pm$0.033 & 0.01 \\
Ent. (mid) & 0.680$\pm$0.024 & 0.928$\pm$0.050 & 0.915$\pm$0.031 & 0.39 \\
Ent. (small) & 0.683$\pm$0.016 & 0.955$\pm$0.030 & 0.959$\pm$0.036 & 0.40 \\
Degree & 0.685$\pm$0.020 & 0.939$\pm$0.037 & 0.930$\pm$0.039 & 0.01 \\
Vertex ent. & 0.748$\pm$0.121 & 0.926$\pm$0.045 & 0.934$\pm$0.019 & 0.64 \\
Betweenness & 0.768$\pm$0.063 & 0.879$\pm$0.042 & 0.945$\pm$0.022 & 0.04 \\
Eigenvector & 0.934$\pm$0.028 & 0.958$\pm$0.020 & 0.964$\pm$0.023 & 0.01 \\
Random & 0.960$\pm$0.024 & 0.948$\pm$0.020 & 0.964$\pm$0.010 & 0.01 \\
\bottomrule
\end{tabular*}

\end{table}

\begin{figure*}[t]
\centering
\includegraphics[width=\textwidth]{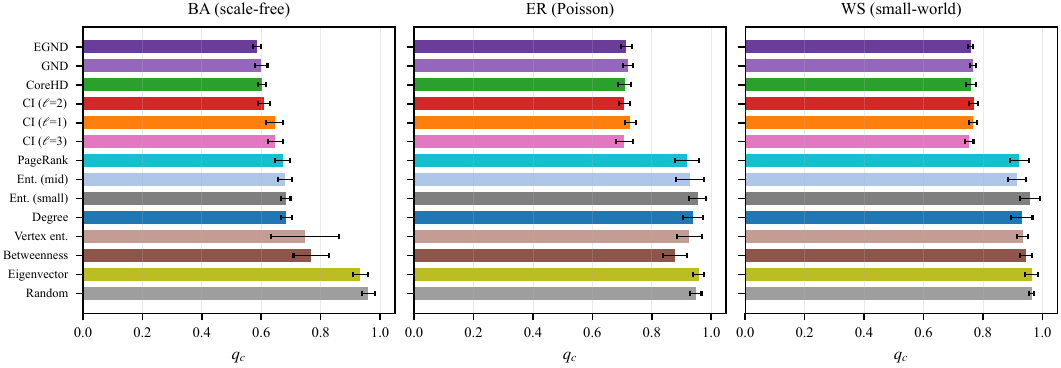}
\caption{Threshold cost at $N=100$ for every procedure measured, grouped by ensemble.
Error bars are one standard deviation over ten realisations. The ordering is stable across
ensembles for the extremes (spectral and collective-influence methods best, eigenvector and
random worst) and unstable in the middle, where entanglement-based and centrality-based
scores interleave within one standard deviation.}
\label{fig:zoo}
\end{figure*}

Table~\ref{tab:zoo} and Fig.~\ref{fig:zoo} widen the comparison to the full measured set.
Three observations.

First, the ensemble average EGND gives the lowest cost on BA ($0.585$) and ties CoreHD on
WS ($0.759$); its advantage over a single GND run is $0.015$ on BA for a tenfold cost
increase. Whether that is worth paying is an engineering decision, but it should be made
explicitly rather than by default.

Second, the entanglement-based scores occupy the middle of the ranking: on BA they are
competitive with PageRank ($0.680$ versus $0.672$) and clearly better than eigenvector, and
on ER and WS they are indistinguishable from degree ordering within one standard deviation.
Their computational cost, however, is two orders of magnitude higher ($0.39$--$0.64$~s
versus $0.01$~s at $N=100$), and it grows faster, so they are not a practical substitute
for CI at larger sizes. Their interest is conceptual---they rank by a spectral
\emph{loss} rather than by a structural proxy---and they would be worth revisiting if a
faster implementation existed.

Third, the middle of the ranking is not resolvable at this sample size. Differences of
$0.01$--$0.03$ in $q_c$ with standard deviations of $0.02$--$0.05$ are not
distinguishable with ten realisations, and we do not claim an ordering there. This is a
statement about the field as much as about our data: published comparisons of dismantling
heuristics routinely report differences of this magnitude without replication, and those
differences should be treated as unresolved.

\subsection{Real Backbone Topologies}

\begin{table*}[t]
\centering
\caption{Eleven real backbone topologies, ordered by size: descriptors, threshold cost for
three procedures, and the ratio $q_c^{\mathrm{CI}}/q_c^{\mathrm{rnd}}$. Source:
\texttt{data/descriptors.csv}, \texttt{data/real\_results.csv}.}
\label{tab:real}
\footnotesize
\begin{tabular*}{\textwidth}{@{\extracolsep{\fill}}l*{11}{r}@{}}
\toprule
Network & $N$ & $M$ & $\langle k\rangle$ & CV$(k)$ & $C$ & $r$ & $\lambda_2$ &
$q_c^{\mathrm{rnd}}$ & $q_c^{\mathrm{deg}}$ & $q_c^{\mathrm{CI}}$ & ratio \\
Aarnet & 19 & 24 & 2.53 & 0.39 & 0.07 & +0.03 & 0.12 & 0.68 & 0.84 & 0.53 & 0.77 \\
Abvt & 23 & 31 & 2.70 & 0.39 & 0.15 & -0.28 & 0.18 & 0.74 & 0.87 & 0.52 & 0.71 \\
Agis & 25 & 30 & 2.40 & 0.63 & 0.05 & -0.51 & 0.21 & 0.72 & 0.56 & 0.56 & 0.78 \\
Geant2009 & 34 & 52 & 3.06 & 0.57 & 0.24 & -0.17 & 0.21 & 0.91 & 0.85 & 0.53 & 0.58 \\
Geant2012 & 40 & 61 & 3.05 & 0.63 & 0.15 & -0.20 & 0.14 & 0.95 & 0.78 & 0.50 & 0.53 \\
Uunet & 49 & 84 & 3.43 & 0.83 & 0.14 & +0.05 & 0.20 & 0.90 & 0.63 & 0.49 & 0.55 \\
Dfn & 58 & 87 & 3.00 & 0.90 & 0.14 & -0.11 & 0.25 & 0.95 & 0.76 & 0.50 & 0.53 \\
Switch & 74 & 92 & 2.49 & 0.62 & 0.11 & -0.37 & 0.04 & 0.81 & 0.73 & 0.54 & 0.67 \\
Uninett2010 & 74 & 101 & 2.73 & 0.57 & 0.03 & +0.14 & 0.07 & 0.88 & 0.77 & 0.54 & 0.62 \\
Pern & 127 & 129 & 2.03 & 2.10 & 0.00 & -0.67 & 0.02 & 0.80 & 0.08 & 0.08 & 0.10 \\
DialtelecomCz & 138 & 151 & 2.19 & 0.39 & 0.02 & -0.11 & 0.01 & 0.86 & 0.81 & 0.48 & 0.56 \\
\bottomrule
\end{tabular*}

\end{table*}

\begin{figure*}[t]
\centering
\includegraphics[width=\textwidth]{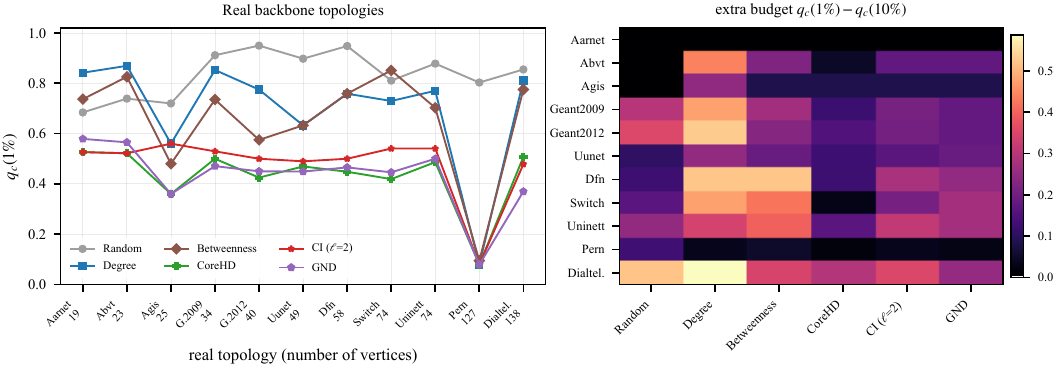}
\caption{Left: threshold cost $q_c(1\%)$ of six procedures across the real topologies; the
second line of each tick label gives the number of vertices. Right: extra budget
$q_c(1\%)-q_c(10\%)$, i.e.\ how many further removals are needed to shrink the largest
component from $10\%$ to $1\%$ of $N$. A small value means the collapse is abrupt: the
second threshold follows the first almost immediately.}
\label{fig:real}
\end{figure*}

Table~\ref{tab:real} reports the eleven backbone topologies. The dominant pattern is
\emph{uniformity of normalised cost}: for ten of the eleven graphs, CI needs between
$0.48$ and $0.56$ of the vertices, despite a sevenfold range in $N$ and a wide range of
descriptors. Only Pern departs from this, and it departs completely: with $N=127$ and
$M=129$ it is essentially a tree, so $q_c^{\mathrm{CI}}=0.08$---eight percent of the
vertices suffice. The lesson is that the normalised worst-case cost of a well-connected
backbone is roughly one half, and that a topology can look unremarkable on every
descriptor ($\langle k\rangle=2.03$, modest clustering) while being an order of magnitude
easier to dismantle.

The right panel of Fig.~\ref{fig:real} adds a second reading. On most backbones the gap
between the $10\%$ and $1\%$ thresholds is large, meaning that after the structure breaks
the remaining fragments are still sizeable and further removals are needed; on Pern and on
Aarnet the gap is small, meaning the collapse is essentially complete at the first
threshold. Two topologies with the same $q_c$ can therefore differ in whether the failure
is abrupt or gradual---which is the real-topology analogue of the $S_2$ diagnostic in
Fig.~\ref{fig:curves}.

Random removal, for comparison, costs $0.68$--$0.95$ on the same graphs, so the
targeted-to-random ratio is $0.10$ (Pern), $0.53$--$0.67$ (most backbones), and $0.78$
(Agis). A defender who budgets for random failures will therefore systematically
underestimate the required redundancy by a factor of $1.5$--$2$, and by a factor of $10$ on
tree-like backbones.

\subsection{Which Descriptors Predict Difficulty?}

\begin{table}[t]
\centering
\caption{Correlation between topology descriptors and dismantling cost $q_c$ (CI, $\ell=2$,
$N=200$). ``Pooled'' uses all $30$ graphs; ``within-family'' averages the Pearson
correlation computed separately inside BA, ER and WS. A descriptor that predicts across
ensembles need not predict within one.}
\label{tab:desccorr}
\footnotesize
\begin{tabular*}{\columnwidth}{@{\extracolsep{\fill}}lcccc@{}}
\toprule
Descriptor & \multicolumn{3}{c}{pooled ($n{=}30$)} & within-family \\
\cmidrule(lr){2-4}
 & $r$ & $\rho$ & $p$ & mean $r$ \\
$\langle k\rangle$ & +0.318 & +0.544 & 1.9e-03 & -0.124 \\
CV$(k)$ & -0.953 & -0.929 & 1.4e-13 & -0.251 \\
$k_{\max}$ & -0.906 & -0.886 & 7.6e-11 & -0.254 \\
$r$ (assort.) & +0.686 & +0.596 & 5.1e-04 & +0.070 \\
$C$ (clustering) & +0.568 & +0.416 & 2.2e-02 & -0.268 \\
$\lambda_2$ & -0.607 & -0.474 & 8.2e-03 & +0.228 \\
$\bar{d}$ (avg. path) & +0.918 & +0.911 & 2.8e-12 & +0.291 \\
diameter & +0.717 & +0.679 & 3.7e-05 & -0.236 \\
max betweenness & -0.855 & -0.713 & 9.8e-06 & -0.337 \\
$k$-core & +0.859 & +0.803 & 9.5e-08 & -0.176 \\
\bottomrule
\end{tabular*}

\end{table}

\begin{figure*}[t]
\centering
\includegraphics[width=\textwidth]{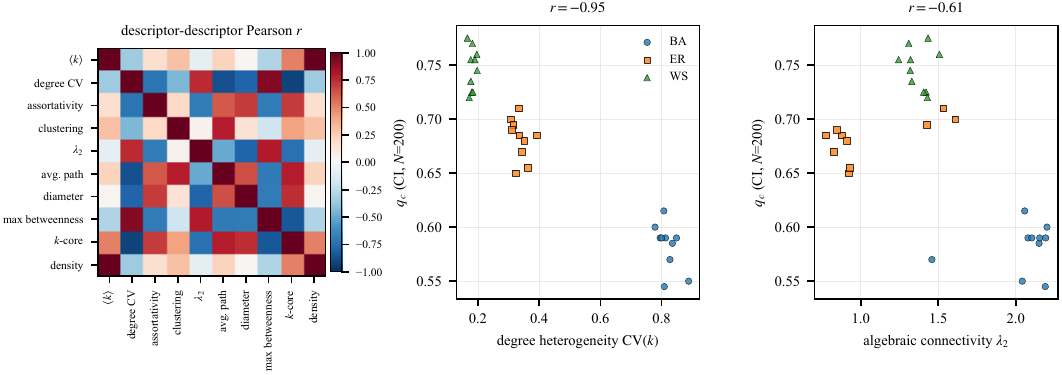}
\caption{Left: descriptor--descriptor correlations, showing that the descriptors are not
independent measurements. Centre and right: dismantling cost against degree heterogeneity
and against algebraic connectivity, coloured by family. The two descriptors separate the
families (different clusters) but do not order the graphs inside a family (overlapping
points).}
\label{fig:descriptors}
\end{figure*}

Table~\ref{tab:desccorr} answers the question the tutorial poses in Step 3, and the answer
is more subtle than the pooled correlations suggest.

Across all $30$ graphs, degree heterogeneity is an almost perfect predictor
($r=-0.953$), followed by average path length ($+0.918$), maximum degree ($-0.906$),
maximum betweenness ($-0.855$) and degeneracy ($+0.859$). Algebraic connectivity, the
metric most often recommended for structural robustness, is a distant sixth ($-0.607$).
Read naively, this says: compute $\mathrm{CV}(k)$ and you know the answer.

The within-family column destroys that reading. Inside a family, the same correlation
collapses to $-0.25$ on average, and for algebraic connectivity it changes sign
($+0.23$). The pooled correlation is therefore mostly a \emph{between-family} effect: it
measures the fact that BA graphs are heterogeneous and hard to break, and WS graphs are
homogeneous and easy to break, which was never in doubt. It does not tell a designer which
of two scale-free topologies is more robust. That question requires measurement, and the
residual variance after conditioning on family is exactly the part that the descriptors
cannot see.

Fig.~\ref{fig:descriptors} shows why. The families form distinct clusters in
$(\mathrm{CV}(k), q_c)$ space, and a single regression line through all clusters has a
steep slope that is driven by the separation between clusters, not by the spread within
them. Reporting the pooled $r$ without the within-family $r$ is a regression instance of
error~8 in Table~\ref{tab:checklist}: attributing to a descriptor an effect that belongs to
the ensemble.

\subsection{Checking the Random Baseline Against Theory}

\begin{figure*}[t]
\centering
\includegraphics[width=\textwidth]{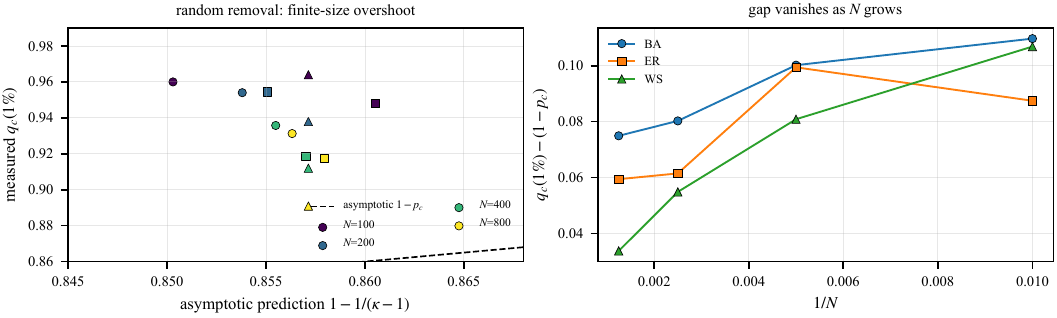}
\caption{Left: measured random-removal threshold against the asymptotic prediction
$1-1/(\kappa-1)$; marker shape encodes the family and colour encodes $N$. Every point lies
above the diagonal, and the vertical distance shrinks with $N$. Right: the same distance
as a function of $1/N$. The residual offset is the finite-size cost of the $1\%$ criterion,
not a failure of the theory.}
\label{fig:percolation}
\end{figure*}

As a check on the whole pipeline---graph generation, ordering, replay, threshold
detection---we compare the measured random-removal threshold with the mean-field prediction
\eqref{eq:pc}. The mean degree is matched at $\langle k\rangle\approx8$ by construction, so
the prediction is essentially the same for all three families
($1-1/(\kappa-1)=0.850$--$0.861$).

Every measured value lies above the prediction: by $0.087$--$0.110$ at $N=100$ and by
$0.034$--$0.075$ at $N=800$, and the excess decreases monotonically in $1/N$ in every
ensemble (Fig.~\ref{fig:percolation}). This is expected and is not evidence against
\eqref{eq:pc}: at $p_c$ the largest component is subextensive but not of size one, so
requiring $|C_{\max}|\le0.01N$ necessarily demands more removals than the percolation
threshold, and the discrepancy is larger when $N$ is small enough for the critical window
to overlap the $1\%$ scale. The check therefore validates the implementation and
simultaneously quantifies the finite-size bias discussed in Section~\ref{sec:eval}---which
is the reason we report $q_c(10\%)$ alongside $q_c(1\%)$ throughout.

\subsection{Summary of Empirical Findings}

\begin{enumerate}
\item \textbf{Trajectory shape, not just cost, distinguishes ensembles.} BA erodes, ER
collapses with a pronounced $S_2$ peak, WS resists local attack and yields only to global
procedures.
\item \textbf{The targeted--random gap is set by heterogeneity and grows with $N$ on
heterogeneous ensembles} ($0.275\to0.455$ on BA) while remaining small on homogeneous ones
($0.056$ on WS at $N=800$ for degree ordering).
\item \textbf{CI is the most consistent heuristic} across ensembles and sizes, at the
lowest computational cost; \textbf{GND is competitive only for $N\lesssim400$}, where its
exact vertex cover pays off.
\item \textbf{Two popular centralities fail:} betweenness is dominated by degree on two of
three ensembles and is the most expensive procedure, and eigenvector centrality is worse
than random removal on the small-world ensemble.
\item \textbf{Area and threshold can invert a ranking} for specific method pairs
(GND versus CoreHD at $N=800$), so the summary statistic must be named.
\item \textbf{Real backbones have a near-universal normalised worst-case cost} of about one
half of their vertices, with tree-like outliers an order of magnitude cheaper.
\item \textbf{Descriptors predict across ensembles but not within them:} the pooled
$r=-0.95$ of degree heterogeneity falls to $-0.25$ inside a family, and algebraic
connectivity changes sign.
\end{enumerate}

\section{Tutorial: From a Topology to Defensible Evidence}
\label{sec:tutorial}

The preceding sections define metrics and methods. This section turns them into a
procedure that produces evidence a reader can audit. It is written for the practitioner
who has one topology---a planned constellation, a backbone upgrade, a
software-defined network snapshot---and needs to state something true about its structural
margin. The procedure has seven steps; each produces an artefact that the next step
consumes.

\subsection{Step 1: Ask a Question the Graph Can Answer}

Begin with a question whose answer is a property of the graph under a stated disturbance.
Three forms are useful, in increasing order of strength:

\begin{enumerate}
\item \emph{Descriptive with a baseline}: ``how many vertex removals does this topology
require before no fragment exceeds $10\%$ of the nodes?''
\item \emph{Comparative}: ``does this topology require more removals than a random graph
with the same degree sequence?''
\item \emph{Adversarial}: ``what is the smallest number of removals any of the tested
procedures needs?''
\end{enumerate}

The first form is a measurement; the second is a controlled comparison and requires the
null model to be stated; the third is a bound and requires the procedure set to be stated.
A question that mixes them---``is this network resilient?''---cannot be answered, and
attempting to answer it is the root cause of most of the errors in
Table~\ref{tab:checklist}.

\subsection{Step 2: Declare the Graph}

Record, before any computation: what a vertex is (router, satellite, PoP, city), what an
edge is (physical fibre, a feasible line-of-sight link, a measured adjacency), and under
which operating condition the edge set was taken. Then record four decisions:

\begin{itemize}
\item \textbf{Isolated vertices.} Keep a node roster and add isolated devices explicitly;
a topology file that lists only edges silently drops them, and dropping isolated vertices
inflates every connectivity metric.
\item \textbf{Disconnected input.} If the stored graph is disconnected, either evaluate it
as is or extract the largest component---but state which. Extraction changes the evaluated
deployment and makes $N$ smaller, which changes $q_c(1\%)$ through the finite-size effect
of Section~\ref{sec:eval}.
\item \textbf{Directed and weighted data.} Symmetrisation and binarisation are
irreversible; record that they were applied.
\item \textbf{Identification.} Preserve original identifiers so that a removal order can
be reported in the operator's own vocabulary.
\end{itemize}

\subsection{Step 3: Pre-Screen with Descriptors}

Descriptors cost seconds and eliminate whole classes of questions. Compute
$\langle k\rangle$, $\mathrm{CV}(k)$, assortativity $r$, clustering $C$, algebraic
connectivity $\lambda_2$, and degeneracy. Two readings are immediately available:

\begin{itemize}
\item From \eqref{eq:pc}, the random-removal threshold is approximately
$1-1/(\kappa-1)$ with $\kappa=\langle k^2\rangle/\langle k\rangle$. A topology with
$\kappa\approx3$ cannot survive more than about half of its vertices failing at random,
regardless of how it is wired; if the design requirement is ``tolerate $50\%$ random
loss'', the ensemble is infeasible before any simulation.
\item A large $\mathrm{CV}(k)$ together with a small $\lambda_2$ is the signature of a
hub-and-spoke topology with weak lateral connectivity: robust to random loss, fragile to
targeted loss. This prediction is testable in Step 5 and, if it fails, is informative.
\end{itemize}

Descriptors are a filter, not a verdict. Section~\ref{sec:results} shows that they explain
a substantial share of the variance in dismantling cost but not all of it, and that the
residual is where the topology-specific structure lives.

\subsection{Step 4: Choose the Protocol}

Run at least three procedures, always including random as a null:

\begin{enumerate}
\item \textbf{Random} (seeded, several seeds): the null model. Without it, a targeted
result cannot be interpreted.
\item \textbf{One cheap targeted procedure}: degree ordering is the standard choice, and
it is also the one an unsophisticated adversary would use.
\item \textbf{One optimised procedure}: CI ($\ell=2$) for large graphs, GND for graphs
below a few thousand vertices where its cost is acceptable.
\end{enumerate}

If the three answers differ by less than the realisation noise, the topology is
\emph{structurally homogeneous} and the choice of attacker does not matter---a useful
conclusion in itself. If they differ substantially, the gap \emph{is} the result, and it
should be reported as the headline number.

\subsection{Step 5: Run, Record, and Replay}

The following fragment is the complete measurement for one topology. It uses the harness
of this paper, which in turn calls the library's public interface.

\begingroup\scriptsize
\begin{verbatim}
from pathlib import Path
import networkx as nx
from bench.common import (normalize, run_method,
                          topology_descriptors)

G = nx.read_edgelist("topology.edgelist",
                     data=False, nodetype=str)
G.add_nodes_from(
    Path("nodes.txt").read_text().split())
G = normalize(G)      # undirected, simple, 0..n-1
print(topology_descriptors(G))

for method in ("random", "degree",
               "CI_L2", "GND"):
    r = run_method(G, method, seed=0)
    print(f"{method:10s} q_c={r['q_c']:.3f} "
          f"q_c10={r['q_c_10']:.3f} "
          f"A={r['auc']:.3f} "
          f"maxS2={r['slcc_peak']:.3f} "
          f"t={r['time']:.2f}s")
\end{verbatim}
\endgroup

Three details make the output reproducible. The seed is explicit for every procedure that
has one. The full trajectory is retained, not only the summary, so that a different
threshold can be read later without re-running. And the target is read from the native
prefix (Step 2 of Section~\ref{sec:eval}), so changing $\tau$ does not change the
procedure.

\subsection{Step 6: Verify Before Interpreting}

Two checks catch most implementation errors, and both are cheap.

\textbf{An analytical case.} Take the $5$-vertex star and remove the hub first. The
trajectory is $S=(1,0.2,0.2,0.2,0.2,0)$ over $q\in\{0,0.2,\dots,1\}$, hence
\begin{equation}
A=\tfrac15\Big[\tfrac{1+0.2}{2}+3\cdot\tfrac{0.2+0.2}{2}+\tfrac{0.2+0}{2}\Big]=0.26 ,
\end{equation}
and the initial global efficiency \eqref{eq:eff} is
$\frac{1}{20}\left(8\cdot1+12\cdot\frac12\right)=0.7$. An evaluator that does not
reproduce both numbers has a normalisation or an endpoint bug.

\textbf{A structural case.} Recompute the trajectory by \emph{reverse addition}: start
from the empty graph and add vertices back in reverse removal order, maintaining component
sizes with a union--find structure. The two trajectories must agree exactly. This check is
independent of the forward implementation and catches off-by-one errors at the sequence
boundary, which are otherwise invisible except as a small bias in $q_c$.

\subsection{Step 7: Report}

A report that can be audited contains, at minimum: the graph declaration (Step 2); the
descriptor table (Step 3); the protocol table, listing each procedure, its adaptivity, its
native stopping rule, and the seed set (Step 4); the trajectory figure with $S$ and $S_2$;
the threshold table for at least two values of $\tau$; and the runtime, measured on the
same machine as the comparison. Table~\ref{tab:checklist} closes this section with the
errors that most often survive into publication.

\subsection{Transferring the Procedure}

To apply the procedure to a new topology, only Step 2 changes. Supply an undirected
two-column edge list plus a node roster, decide the disconnected-input treatment, and run
the same fragment. For sizes at which efficiency \eqref{eq:eff} is affordable, add it as a
second observable: it separates ``the network still connects'' from ``the network still
connects \emph{well}'', and the two can diverge sharply---a point we quantify in
Section~\ref{sec:results}.

\begin{table*}[t]
\caption{Nine recurring interpretation errors, the check that detects each, and the
section that specifies the fix.}
\label{tab:checklist}
\centering
\footnotesize
\begin{tabularx}{\textwidth}{@{}p{0.5cm} L L p{1.3cm}@{}}
\toprule
\# & Error & Check & Fix \\
\midrule
1 & Reporting a metric without its disturbance protocol & Can a reader reproduce the removal order from the text? & Sec.~\ref{sec:eval} \\
2 & Comparing static and adaptive procedures as if equivalent & Is the adaptivity flag recorded per row? & Tab.~\ref{tab:methods} \\
3 & Reading $q_c$ from the completed sequence rather than the native prefix & Do $q_c$ values shift when $\tau$ changes? & Alg.~\ref{alg:eval} \\
4 & Using $N_{\text{surviving}}$ as the denominator of $S$ & Does a 3-node fragment report $S=1$? & Sec.~\ref{sec:metrics} \\
5 & Treating $A$ and $q_c$ as interchangeable summaries & Do the two rankings agree on your data? & Sec.~\ref{sec:results} \\
6 & Comparing $q_c(1\%)$ across sizes without noting the finite-size effect & At $N{=}100$, is $q_c$ close to 1 for every method? & Sec.~\ref{sec:eval} \\
7 & Counting thresholds as independent repetitions & Are $\tau$ values read from one trajectory? & Sec.~\ref{sec:eval} \\
8 & Attributing a ranking to the algorithm when it is an ensemble artefact & Does the ranking survive a change of family or $N$? & Sec.~\ref{sec:results} \\
9 & Extrapolating a structural result to service performance & Is a flow or capacity model stated? & Sec.~\ref{sec:discussion} \\
\bottomrule
\end{tabularx}
\end{table*}

\section{Discussion, Open Problems, and Conclusion}
\label{sec:discussion}

\subsection{Limitations of This Study}

The benchmark of Section~\ref{sec:results} is deliberately narrow, and three of its
boundaries should be read as scoping decisions rather than oversights.

\emph{Unweighted, undirected, vertex-only.} Capacity, delay, and direction change both the
connectivity predicate and the cost of the algorithms. The protocol transfers, but the
constants do not; a weighted network in which links carry heterogeneous capacity can be
structurally intact and functionally destroyed, which is precisely the gap that
distance-based metrics of Section~\ref{sec:metrics} are designed to expose.

\emph{Finite size and finite replication.} Ten realisations at four sizes up to $N=800$
suffice to separate methods that differ by more than the realisation noise, and we have
avoided claims about differences smaller than that. They do not suffice to establish
asymptotic rankings, and the \emph{reversal} we observe between $N=100$ and $N=800$ for
the two centrality rules is a warning against exactly such extrapolation.

\emph{Incomplete method coverage.} GDM, FINDER, NIRM, and AutoSND are surveyed but not
measured, for the packaging and environment reasons documented in
Section~\ref{sec:library}. This matters for the claim ``which method is best'': the
learned methods would plausibly occupy the top of the ranking at small $N$, and their
absence makes our ranking a ranking of \emph{classical and spectral} methods. We state
this in the caption of every table rather than in a footnote.

\subsection{Relation to the Wider Resilience Literature}

``Resilience'' entered the quantitative literature through ecology, as the capacity of a
system to absorb disturbance and reorganise while retaining function
\cite{holling1973resilience}; engineering and regional-economics traditions subsequently
sharpened it into a time-domain property with an absorption, a recovery, and an adaptation
phase \cite{proag2014concept, martin2015regional, bergstrom2015rationale,
wied2020conceptualizing, haimes2008homeland}. Every measure in that tradition is a
functional of a performance curve over \emph{time}, not over \emph{removal budget}. The
distinction matters: our $S(q)$ has the same shape as a resilience curve and is often
presented as one, but the abscissa is an attack budget, and the recovery limb is absent.

The bridge between the two is the observation that a structural trajectory provides the
\emph{input} to a time-domain model. If one knows how performance falls as a function of
damage, and one has a model of how damage accumulates in time (failure arrivals, repair
rates, cascade dynamics), then the time-domain resilience integral follows. Analytical
efforts to make that link quantitative include universal resilience patterns derived by
dimension reduction \cite{gao2016universal}, collapse prediction from early-warning
signals \cite{wu2024collapse, duan2025resilience, panahi2024machine}, and learned
resilience inference \cite{liu2024resinf}; reviews of the assessment landscape are given in
\cite{cassottana2023resilience, qi2024network, fraccascia2018resilience,
XU2024tsosra}. Importance-measure approaches \cite{Dui2023ImportanceMR} occupy an
intermediate position: they rank components by their contribution to a resilience
functional, which is the time-domain analogue of the vertex rankings studied here.

For coupled systems the picture is further complicated by inter-layer propagation
\cite{battiston2020networks, mahabadi2021network, danziger2022recovery,
ferrazdearruda2024contagion}, where a structural removal in one layer can trigger
functional failure in another. The multilayer dismantling literature
\cite{Gu2025DeeplearningaidedDO, Osat2017OptimalPO, hypergraphdismantling2026} is the
structural counterpart.

\subsection{Applications of the Dismantling Viewpoint}

The same computation that bounds worst-case damage also answers questions that are not
adversarial. In epidemic control, the removal set is a vaccination or quarantine target,
and the percolation formulation is exactly the one used for immunization strategies
\cite{pastor2001epidemic, boguna2013nature, kinouchi2006optimal}; the explosive-percolation
heuristic behind EI was proposed in that context \cite{Clusella2016ImmunizationAT,
achlioptas2009explosive}. In infrastructure protection, the removal set identifies assets
whose duplication removes a bottleneck, and network isolators exploit the same cut
structure defensively \cite{kaiser2024network}; flow-weighted variants extend the idea to
networks where connectivity is not the binding constraint \cite{zhu2025small,
zhu2025can}. In systems biology the targets are proteins or metabolites, and the objective
is a functional one rather than a connectivity one, which is why the survey stops at the
structural boundary.

The complementary design problem---\emph{changing} the topology to raise its dismantling
cost---is deliberately outside our scope, but it is where most of the recent engineering
literature lives: reinforcement-learning rewiring \cite{li2023network, yang2023learning},
IoT-specific robustness evolution \cite{peng2022smarttro, chen2024daimo, qiu2023self},
multi-objective and memetic construction \cite{Wang2018AME, Zhou2014AMA}, simulated
annealing and cascade-aware optimisation \cite{Buesser2011OptimizingTR,
gutfraind2011optimizing, herrmannonionlike2011}, and criticality-based dynamic
optimisation for swarm networks \cite{swarmtopo2025}. A reader who has completed the
evaluation of this paper is in a position to use those methods correctly: the evaluation
supplies the objective function, and the redesign literature supplies the search.

\subsection{Open Problems}

\textbf{Reporting standards.} The single most valuable contribution the community could
make is a mandatory protocol block in every dismantling paper: graph declaration, target
definition, adaptivity, native stopping rule, seed set, and runtime environment. The
recent benchmarking efforts \cite{netbreaker2026, Wandelt2018ACA} move in this direction;
the present paper is an argument that the block should be standard, because without it
numbers are not comparable even when the algorithms are.

\textbf{Evaluation at operational scale.} Full-trajectory evaluation costs
$O(N(N+M))$, which is prohibitive for $N\gtrsim10^6$ in interpreted code. Two directions
are plausible: incremental component maintenance under deletions (the reverse-addition
union--find trick used as a verification in Section~\ref{sec:tutorial} generalises to a
forward structure with rollback), and \emph{online} evaluation in which the metric is
updated as the topology changes rather than recomputed. The second is what a deployed
network actually needs, and it is largely open.

\textbf{Temporal topologies.} Satellite constellations, UAV swarms, and tactical edge
networks have topologies that change on a timescale comparable to the attack. The
structural trajectory $S(q)$ then becomes a family $S(q,t)$, and the relevant question is
not ``how many removals'' but ``how many removals \emph{within one connectivity window}''
\cite{farhan2025rdg, digitaltwinleo2026, holme2012temporal}. Temporal percolation is
developing rapidly \cite{quantumperc2026, smallnetworks2025} but has not produced a
consensus evaluation protocol.

\textbf{Higher-order and multilayer structure.} Hypergraph and multiplex dismantling
\cite{hypergraphdismantling2026, Gu2025DeeplearningaidedDO, Osat2017OptimalPO,
higherordercascade2025} change the objective rather than the solver: in a hypergraph, a
single removal can destroy many pairwise connections, so the ``cost'' of a removal is no
longer one. Whether the classical summaries ($q_c$, $A$) remain the right ones in this
setting is unresolved.

\textbf{Learned methods and their evidence.} Three specific questions recur. First,
\emph{leakage}: training on graphs from the same generator as the test set inflates
performance, and the field has no standard split. Second, \emph{transfer}: a policy trained
on synthetic ensembles is applied to backbones with qualitatively different structure
(planar, clustered, low degree), and the degradation is rarely quantified
\cite{nohandcraft2025}. Third, \emph{reproducibility}: checkpoints, preprocessing, and
inference environments are often unavailable, which is why a library that pins them is more
valuable than a leaderboard.

\textbf{Theory of the finite-size regime.} Universality classes for dismantling
transitions have been characterised asymptotically \cite{criticality2026}, but engineering
decisions are made at $N=10^2$--$10^3$. A finite-size scaling theory that predicts
$q_c(N)$ from descriptors would replace the empirical pre-screening of
Section~\ref{sec:tutorial} with a formula---and would make the descriptor regressions of
Section~\ref{sec:results} interpretable rather than merely predictive.

\subsection{Dual-Use Considerations}

Dismantling algorithms are attack algorithms. The same procedure that identifies the
vertices an adversary should remove identifies the vertices a defender should protect,
duplicate, or monitor, and the literature is explicit that the defensive use is the
intended one \cite{Schneider2011MitigationOM, Morone2015InfluenceMI}. Two practical
consequences follow. First, rankings of \emph{critical} vertices are sensitive and should
be handled as such when they concern deployed infrastructure. Second, the marginal risk of
publishing a heuristic is low---the methods are already public and reproducible---whereas
the marginal benefit of publishing a \emph{protocol} is high, because protocol standardisation
is what allows a defender to know whether a claimed robustness figure is real.

\subsection{Conclusion}

This paper treated two questions and their interdependence: how the quality of a network
topology should be measured, and how quickly it can be broken. The first answer is that a
structural metric is a functional of a disturbance protocol, so metrics must be indexed by
the disturbance they presuppose; the second is that dismantling algorithms are lower-bound
generators whose output is a curve, not a number.

On that basis we organised the metric literature by observable and the algorithmic
literature by objective, implemented twenty-two methods behind a single interface,
measured ten of them across four sizes and fourteen at one size, on three synthetic
ensembles and eleven real backbones with a paired, replicated protocol, and converted the
result into a tutorial with two verification checks and a nine-item error checklist. The
empirical findings that we consider most portable are these: aggregate area and threshold
cost agree in the aggregate yet invert specific method pairs, so the summary statistic must
be named; the targeted-versus-random gap tracks degree heterogeneity and \emph{widens}
with $N$ on heterogeneous ensembles while staying small on small-world ones; CI is the most
consistent heuristic, whereas eigenvector centrality is no better than random removal and
betweenness is dominated on both effectiveness and cost; and pooled descriptor correlations
overstate predictive power, since degree heterogeneity explains $95\%$ of the variance
across ensembles but only $25\%$ within one, which makes descriptors a screening tool and
not a substitute for measurement.

The broader point is methodological. In a field where every paper proposes a new
procedure, the durable contribution is not the procedure but the conditions under which it
can be compared. A topology evaluation that states its disturbance, its observable, its
normalisation, and its ensemble is reproducible even if its conclusion is negative; one
that does not is unreproducible even if its conclusion is favourable.

\balance
\bibliographystyle{IEEEtran}
\bibliography{refs}

@article{erdos1960evolution,
  title={On the evolution of random graphs},
  author={Erd{\H{o}}s, Paul and R{\'e}nyi, Alfr{\'e}d},
  journal={Publ. Math. Inst. Hung. Acad. Sci},
  volume={5},
  number={1},
  pages={17--60},
  year={1960}
}

@article{watts1998collective,
  title={Collective dynamics of ‘small-world’ networks},
  author={Watts, Duncan J and Strogatz, Steven H},
  journal={Nature},
  volume={393},
  number={6684},
  pages={440--442},
  year={1998},
  publisher={Nature Publishing Group}
}

@article{barabasi1999emergence,
  title={Emergence of scaling in random networks},
  author={Barab{\'a}si, Albert-L{\'a}szl{\'o} and Albert, R{\'e}ka},
  journal={Science},
  volume={286},
  number={5439},
  pages={509--512},
  year={1999},
  publisher={American Association for the Advancement of Science}
}

@article{albert2000error,
  title={Error and attack tolerance of complex networks},
  author={Albert, R{\'e}ka and Jeong, Hawoong and Barab{\'a}si, Albert-L{\'a}szl{\'o}},
  journal={Nature},
  volume={406},
  number={6794},
  pages={378--382},
  year={2000},
  publisher={Nature Publishing Group}
}

@article{cohen2000resilience,
  title={Resilience of the internet to random breakdowns},
  author={Cohen, Reuven and Erez, Keren and Ben-Avraham, Daniel and et al.},
  journal={Physical review letters},
  volume={85},
  number={21},
  pages={4626},
  year={2000},
  publisher={APS}
}

@article{molloy1995critical,
  title={A critical point for random graphs with a given degree sequence},
  author={Molloy, Michael and Reed, Bruce},
  journal={Random Structures \& Algorithms},
  volume={6},
  number={2-3},
  pages={161--180},
  year={1995},
  publisher={Wiley Online Library}
}

@article{dorogovtsev2008critical,
  title={Critical phenomena in complex networks},
  author={Dorogovtsev, Sergey N and Goltsev, Alexander V and Mendes, Jos{\'e} FF},
  journal={Reviews of Modern Physics},
  volume={80},
  number={4},
  pages={1275--1335},
  year={2008},
  publisher={APS}
}

@book{barabasi2016networkscience,
  author    = {Barab{\'a}si, Albert-L{\'a}szl{\'o}},
  title     = {Network Science},
  year      = {2016},
  publisher = {Cambridge University Press}
}

@article{pastor2001epidemic,
  title={Epidemic spreading in scale-free networks},
  author={Pastor-Satorras, Romualdo and Vespignani, Alessandro},
  journal={Physical review letters},
  volume={86},
  number={14},
  pages={3200},
  year={2001},
  publisher={APS}
}

@article{boguna2013nature,
  title={Nature of the Epidemic Threshold for the Susceptible-Infected-Susceptible Dynamics in Networks},
  author={Bogun{\'a}, Marian and Castellano, Claudio and Pastor-Satorras, Romualdo},
  journal={Physical review letters},
  volume={111},
  number={6},
  pages={068701},
  year={2013},
  publisher={APS}
}

@article{Li2021PercolationOC,
  title={Percolation on complex networks: Theory and application},
  author={Ming Li and Run-Ran Liu and Linyuan L{\"u} and et al.},
  journal={Physics Reports},
  year={2021},
  volume={907},
  pages={1-68}
}

@article{serrano2006percolation,
  title={Percolation and epidemic thresholds in clustered networks},
  author={Serrano, M {\'A}ngeles and Bogun{\'a}, Mari{\'a}n},
  journal={Physical review letters},
  volume={97},
  number={8},
  pages={088701},
  year={2006},
  publisher={APS}
}

@article{rohlf2002criticality,
  title={Criticality in random threshold networks: annealed approximation and beyond},
  author={Rohlf, Thimo and Bornholdt, Stefan},
  journal={Physica A: Statistical Mechanics and its Applications},
  volume={310},
  number={1-2},
  pages={245--259},
  year={2002},
  publisher={Elsevier}
}

@article{rong2022estimation,
  title={On the estimation of percolation thresholds for real networks},
  author={Rong, Qingnan and Zhang, Jun and Sun, Xiaoqian and et al.},
  journal={Chaos, Solitons \& Fractals},
  volume={158},
  pages={111968},
  year={2022},
  publisher={Elsevier}
}

@article{kinouchi2006optimal,
  title={Optimal dynamical range of excitable networks at criticality},
  author={Kinouchi, Osame and Copelli, Mauro},
  journal={Nature physics},
  volume={2},
  number={5},
  pages={348--351},
  year={2006},
  publisher={Nature Publishing Group UK London}
}

@article{Schneider2011MitigationOM,
  title={Mitigation of malicious attacks on networks},
  author={Christian M. Schneider and Andr{\'e} A. Moreira and Jos{\'e} S. Andrade and et al.},
  journal={Proceedings of the National Academy of Sciences},
  year={2011},
  volume={108},
  pages={3838 - 3841}
}

@article{Wu2011SpectralMO,
  title={Spectral Measure of Structural Robustness in Complex Networks},
  author={Jun Wu and Mauricio Barahona and Yuejin Tan and et al.},
  journal={IEEE Transactions on Systems, Man, and Cybernetics},
  year={2011},
  volume={41},
  pages={1244-1252},
}

@article{tutorial_latora2001efficiency,
  author = {Latora, Vito and Marchiori, Massimo},
  title = {Efficient Behavior of Small-World Networks},
  journal = {Physical Review Letters},
  volume = {87},
  pages = {198701},
  year = {2001},
  doi = {10.1103/PhysRevLett.87.198701}
}

@article{Tian2016ArticulationPI,
  title={Articulation points in complex networks},
  author={Liang Tian and Amir Bashan and Da-Ning Shi and et al.},
  journal={Nature Communications},
  year={2016},
  volume={8}
}

@article{engsig2024domirank,
  title={DomiRank Centrality reveals structural fragility of complex networks via node dominance},
  author={Engsig, Mads and Tejedor, Alberto and Moreno, Yamir and et al.},
  journal={Nature Communications},
  volume={15},
  number={1},
  pages={56},
  year={2024}
}

@article{Liu2017ACS,
  title={A comparative study of network robustness measures},
  author={Jing Liu and Mingxing Zhou and Shuai Wang and et al.},
  journal={Frontiers of Computer Science},
  year={2017},
  volume={11},
  pages={568-584}
}

@article{wan2021survey,
  title={A survey on centrality metrics and their network resilience analysis},
  author={Wan, Zelin and Mahajan, Yash and Kang, Beom Woo and et al.},
  journal={IEEE Access},
  volume={9},
  pages={104773--104819},
  year={2021},
  publisher={IEEE}
}

@article{lou2023structural,
  title={Structural robustness of complex networks: A survey of a posteriori measures [feature]},
  author={Lou, Yang and Wang, Lin and Chen, Guanrong},
  journal={IEEE Circuits and Systems Magazine},
  volume={23},
  number={1},
  pages={12--35},
  year={2023},
  publisher={IEEE}
}

@article{artime2024robustness,
  title={Robustness and resilience of complex networks},
  author={Artime, Oriol and Grassia, Marianna and De Domenico, Manlio and et al.},
  journal={Nature Reviews Physics},
  volume={6},
  number={2},
  pages={114--131},
  year={2024}
}

@article{liu2022network,
  title={Network resilience},
  author={Liu, Xueming and Li, Daqing and Ma, Manqing and et al.},
  journal={Physics Reports},
  volume={971},
  pages={1--108},
  year={2022},
  publisher={Elsevier}
}

@article{cerqueti2019measuring,
  title={Measuring network resilience through connection patterns},
  author={Cerqueti, Roy and Ferraro, Giovanna and Iovanella, Antonio},
  journal={Reliability Engineering \& System Safety},
  volume={188},
  pages={320--329},
  year={2019},
  publisher={Elsevier}
}

@article{ahmadian2020quantitative,
  title={A quantitative approach for assessment and improvement of network resilience},
  author={Ahmadian, Navid and Lim, Gino J and et al.},
  journal={Reliability Engineering \& System Safety},
  volume={200},
  pages={106977},
  year={2020},
  publisher={Elsevier}
}

@article{zhang2015assessing,
  title={Assessing the role of network topology in transportation network resilience},
  author={Zhang, Xiaodong and Miller-Hooks, Elise and Denny, Kevin},
  journal={Journal of transport geography},
  volume={46},
  pages={35--45},
  year={2015},
  publisher={Elsevier}
}

@article{edwards2024method,
  title={Method for measuring resilience of complex systems using network theory and graph energy},
  author={Edwards, Christine M and Vierlboeck, Maximilian and Nilchiani, Roshanak Rose and et al.},
  journal={IEEE Open Journal of Systems Engineering},
  volume={2},
  pages={15--25},
  year={2024},
  publisher={IEEE}
}

@article{edwards2024impact,
  title={Impact of graph energy on a measurement of resilience for tipping points in complex systems},
  author={Edwards, Christine M and Nilchiani, Roshanak Rose and Miller, Ian M},
  journal={Systems Engineering},
  volume={27},
  number={4},
  pages={745--758},
  year={2024},
  publisher={Wiley Online Library}
}

@article{fraccascia2018resilience,
  title={Resilience of complex systems: State of the art and directions for future research},
  author={Fraccascia, Luca and Giannoccaro, Ilaria and Albino, Vito},
  journal={Complexity},
  volume={2018},
  number={1},
  pages={3421529},
  year={2018},
  publisher={Wiley Online Library}
}

@article{qi2024network,
  title={Network Resilience: Definitions, approaches, and applications},
  author={Qi, Xiaoyu and Mei, Gang},
  journal={Journal of King Saud University-Computer and Information Sciences},
  volume={36},
  number={1},
  pages={101882},
  year={2024},
  publisher={Elsevier}
}

@article{Morone2015InfluenceMI,
  title={Influence maximization in complex networks through optimal percolation},
  author={F. Morone and Hern{\'a}n A. Makse},
  journal={Nature},
  year={2015},
  volume={524},
  pages={65-68}
}

@article{Morone2016CollectiveIA,
  title={Collective Influence Algorithm to find influencers via optimal percolation in massively large social media},
  author={F. Morone and Byungjoon Min and Lin Bo and et al.},
  journal={Scientific Reports},
  year={2016},
  volume={6}
}

@article{tutorial_morone2016ci,
  author = {Morone, Flaviano and Min, Byungjoon and Bo, Lin and Mari, Romain and Makse, Hern{\'a}n A.},
  title = {Collective Influence Algorithm to find influencers via optimal percolation in massively large social media},
  journal = {Scientific Reports},
  volume = {6},
  pages = {30062},
  year = {2016},
  doi = {10.1038/srep30062}
}

@article{Zdeborov2016FastAS,
  title={Fast and simple decycling and dismantling of networks},
  author={Lenka Zdeborov{\'a} and Pan Zhang and Haijun Zhou},
  journal={Scientific Reports},
  year={2016},
  volume={6}
}

@article{Braunstein2016NetworkD,
  title={Network dismantling},
  author={Alfredo Braunstein and Luca Dall’Asta and Guilhem Semerjian and et al.},
  journal={Proceedings of the National Academy of Sciences},
  year={2016},
  volume={113},
  pages={12368 - 12373}
}

@article{Ren2018GeneralizedND,
  title={Generalized network dismantling},
  author={Xiaolong Ren and Niels Gleinig and Dirk Helbing and et al.},
  journal={Proceedings of the National Academy of Sciences of the United States of America},
  year={2018},
  volume={116},
  pages={6554 - 6559}
}

@article{Wandelt2018ACA,
  title={A comparative analysis of approaches to network-dismantling},
  author={Sebastian Wandelt and Xiaoqian Sun and Daozhong Feng and et al.},
  journal={Scientific Reports},
  year={2018},
  volume={8}
}

@article{Wandelt2021FromRF,
  title={From random failures to targeted attacks in network dismantling},
  author={Sebastian Wandelt and Wei Lin and Xiaoqian Sun and et al.},
  journal={Reliability Engineering \& System Safety},
  year={2021},
  volume={218},
  pages={108146}
}

@article{Mugisha2016IdentifyingOT,
  title={Identifying optimal targets of network attack by belief propagation},
  author={Salomon Mugisha and Haijun Zhou},
  journal={Physical review. E},
  year={2016},
  volume={94 1-1},
  pages={012305}
}

@article{Clusella2016ImmunizationAT,
  title={Immunization and targeted destruction of networks using explosive percolation},
  author={Pau Clusella and Peter Grassberger and Francisco J. P{\'e}rez-Reche and et al.},
  journal={Physical review letters},
  year={2016},
  volume={117 20},
  pages={208301}
  }

@article{Feng2023GeneralizedND,
  title={Generalized network dismantling via a novel spectral partition algorithm},
  author={Zhidan Feng and Zhulou Cao and Xingqin Qi},
  journal={Information Sciences},
  year={2023},
  volume={632},
  pages={285-298}
  }

@article{Feng2024ANA,
  title={A novel algorithm for the generalized network dismantling problem based on dynamic programming},
  author={Zhidan Feng and Huimin Song and Xingqin Qi},
  journal={Chaos, Solitons \& Fractals},
  year={2024}
}

@article{lin2022efficient,
  title={Efficient network dismantling through genetic algorithms},
  author={Lin, Wei and Wandelt, Sebastian and Sun, Xiaoqian},
  journal={Soft Computing},
  volume={26},
  number={6},
  pages={3107--3125},
  year={2022},
  publisher={Springer}
}

@article{Liu2022ANC,
  title={A new centrality measure based on neighbor loop structure for network dismantling},
  author={Qingxia Liu and Bang Wang and Jiming Qi and et al.},
  journal={Digital Communications and Networks},
  year={2022},
  volume={10},
  pages={472-480}
}

@article{Li2019ANC,
  title={A New Complex Network Robustness Attack Algorithm},
  author={Xinyi Li and Zijian Zhang and J. Liu and et al.},
  journal={Proceedings of the 2019 ACM International Symposium on Blockchain and Secure Critical Infrastructure},
  year={2019}
}

@article{zareie2018dnc,
  title={A new centrality measure for identifying influential nodes in social networks},
  author={Zareie, Ahmad and Sheikhahmadi, Amir},
  journal={Physica A: Statistical Mechanics and its Applications},
  volume={492},
  pages={345--358},
  year={2018},
  publisher={Elsevier}
}

@article{Wang2018AME,
  title={A Multi-Objective Evolutionary Algorithm for Promoting the Emergence of Cooperation and Controllable Robustness on Directed Networks},
  author={Shuai Wang and Jing Liu},
  journal={IEEE Transactions on Network Science and Engineering},
  year={2018},
  volume={5},
  pages={92-100}
}

@article{Zhou2017ATM,
  title={A Two-Phase Multiobjective Evolutionary Algorithm for Enhancing the Robustness of Scale-Free Networks Against Multiple Malicious Attacks},
  author={Mingxing Zhou and J. Liu},
  journal={IEEE Transactions on Cybernetics},
  year={2017},
  volume={47},
  pages={539-552}
}

@article{Zhou2014AMA,
  title={A memetic algorithm for enhancing the robustness of scale-free networks against malicious attacks},
  author={Mingxing Zhou and J. Liu},
  journal={Physica A-statistical Mechanics and Its Applications},
  year={2014},
  volume={410},
  pages={131-143}
}

@inproceedings{Buesser2011OptimizingTR,
  title={Optimizing the Robustness of Scale-Free Networks with Simulated Annealing},
  author={Pierre Buesser and Fabio Daolio and Marco Tomassini},
  booktitle={International Conference on Adaptive and Natural Computing Algorithms},
  year={2011}
}

@article{herrmannonionlike2011,
  title = {Onion-like Network Topology Enhances Robustness against Malicious Attacks},
  author = {Herrmann, Hans J and Schneider, Christian M and Moreira, André A and et al.},
  date = {2011-01},
  year = {2011},
  journal = {Journal of Statistical Mechanics: Theory and Experiment},
  volume = {2011},
  number = {01},
  pages = {P01027},
  publisher={IOP Publishing}
}

@incollection{gutfraind2011optimizing,
  title={Optimizing network topology for cascade resilience},
  author={Gutfraind, Alexander},
  booktitle={Handbook of Optimization in Complex Networks: Communication and Social Networks},
  pages={37--59},
  year={2011},
  publisher={Springer}
}

@article{new_targeted_attacks,
  author={Holme, Petter and Kim, Beom Jun and Yoon, Chang No and et al.},
  journal={Physical Review E},
  title={Attack vulnerability of complex networks},
  year={2002},
  volume={65},
  number={5},
  pages={056109}
}

@article{new_cascading_topology,
  author={Motter, Adilson E. and Lai, Ying-Cheng},
  journal={Physical Review E},
  title={Cascade-based attacks on complex networks},
  year={2002},
  volume={66},
  number={6},
  pages={065102(R)}
}

@article{Duan2016ACA,
  title={A comparative analysis of network robustness against different link attacks},
  author={Boping Duan and J. Liu and Mingxing Zhou and et al.},
  journal={Physica A-statistical Mechanics and Its Applications},
  year={2016},
  volume={448},
  pages={144-153}
}

@article{Zhang2020TowardME,
  title={Toward More Effective Centrality-Based Attacks on Network Topologies},
  author={Songwei Zhang and Weisheng Si and Tie Qiu and et al.},
  journal={IEEE International Conference on Communications},
  year={2020},
  pages={1-6}
}

@article{Zhao2021OptimizingTM,
  title={Optimizing the Maximum Vertex Coverage Attacks Under Knapsack Constraint},
  author={Tianming Zhao and Weisheng Si and Wei Li and et al.},
  journal={IEEE/ACM Transactions on Networking},
  year={2021},
  volume={29},
  pages={1088-1104}
}

@article{Osat2017OptimalPO,
  title={Optimal percolation on multiplex networks},
  author={Saeed Osat and Ali Faqeeh and Filippo Radicchi},
  journal={Nature Communications},
  year={2017},
  volume={8}
}

@article{Fan2020FindingKP,
  title={Finding key players in complex networks through deep reinforcement learning},
  author={Changjun Fan and Li Zeng and Yizhou Sun and et al.},
  journal={Nature Machine Intelligence},
  year={2020},
  volume={2},
  pages={317 - 324}
}

@article{Grassia2021MachineLD,
  title={Machine learning dismantling and early-warning signals of disintegration in complex systems},
  author={Marco Grassia and Manlio De Domenico and Giuseppe Mangioni},
  journal={Nature Communications},
  year={2021},
  volume={12}
}

@article{Zhang2023DNDDL,
  title={DND: Deep Learning-Based Directed Network Disintegrator},
  author={Wanchang Zhang and Zhongyuan Jiang and Qingsong Yao},
  journal={IEEE Journal on Emerging and Selected Topics in Circuits and Systems},
  year={2023},
  volume={13},
  pages={841-850}
  }

@article{Zhang2023EncodingND,
  title={Encoding Node Diffusion Competence and Role Significance for Network Dismantling},
  author={Jiazheng Zhang and Bang Wang},
  journal={Proceedings of the ACM Web Conference 2023},
  year={2023}
}

@article{Huang2024IdentifyingKP,
  title={Identifying key players in complex networks via network entanglement},
  author={Yiming Huang and Hao Wang and Xiaolong Ren and et al.},
  journal={Communications Physics},
  year={2024},
  volume={7},
  pages={1-11}
}

@article{Wang2024SpatialND,
  title={Spatial network disintegration based on kernel density estimation},
  author={Zhigang Wang and Zhen Su and Ye Deng and et al.},
  journal={Reliability Engineering \& System Safety},
  year={2024},
  volume={245},
  pages={110005}
}

@article{yang2023learning,
  title={Learning to boost resilience of complex networks via neural edge rewiring},
  author={Yang, Shanchao and Kaili, MA and Wang, Baoxiang and et al.},
  journal={Transactions on Machine Learning Research},
  year={2023}
}

@article{li2023network,
  title={Network topology optimization via deep reinforcement learning},
  author={Li, Zichuan and Chen, Yi-Wei and Liu, Fang and et al.},
  journal={IEEE Transactions on Communications},
  volume={71},
  number={5},
  pages={2847--2859},
  year={2023},
  publisher={IEEE}
}

@article{peng2022smarttro,
  title={SmartTRO: Optimizing topology robustness for Internet of Things via deep reinforcement learning with graph convolutional networks},
  author={Peng, Yabin and Liu, Caixia and Liu, Shuxin and Liu, Yuchen and Wu, Yiteng},
  journal={Computer Networks},
  volume={218},
  pages={109385},
  year={2022},
  publisher={Elsevier}
}

@article{Gu2025DeeplearningaidedDO,
  title={Deep-learning-aided dismantling of interdependent networks},
  author={Weiwei Gu and Chen Yang and Lei Li and et al.},
  journal={Nature Machine Intelligence},
  year={2025},
  volume={7},
  pages={1266 - 1277}
}

@article{liu2024resinf,
  title={Deep learning resilience inference for complex networked systems},
  author={Liu, Cheng and Xu, Feng and Gao, Chao and et al.},
  journal={Nature Communications},
  volume={15},
  number={1},
  pages={9203},
  year={2024},
  publisher={Nature Publishing Group}
}

@article{panahi2024machine,
  title={Machine learning prediction of tipping in complex dynamical systems},
  author={Panahi, Shirin and Kong, Ling-Wei and Moradi, Mohammadamin and et al.},
  journal={Physical Review Research},
  volume={6},
  number={4},
  pages={043194},
  year={2024},
  publisher={APS}
}

@article{kivela2014multilayer,
  title={Multilayer networks},
  author={Kivel{\"a}, Mikko and Arenas, Alex and Barthelemy, Marc and et al.},
  journal={Journal of Complex Networks},
  volume={2},
  number={3},
  pages={203--271},
  year={2014},
  publisher={Oxford University Press}
}

@article{battiston2020networks,
  title={Networks beyond pairwise interactions: structure and dynamics},
  author={Battiston, Federico and Cencetti, Giulia and Iacopini, Iacopo and et al.},
  journal={Physics Reports},
  volume={874},
  pages={1--92},
  year={2020},
  publisher={Elsevier}
}

@article{holme2012temporal,
  title={Temporal networks},
  author={Holme, Petter and Saram{\"a}ki, Jari},
  journal={Physics Reports},
  volume={519},
  number={3},
  pages={97--125},
  year={2012},
  publisher={Elsevier}
}

@article{buldyrev2010catastrophic,
  title={Catastrophic cascade of failures in interdependent networks},
  author={Buldyrev, Sergey V. and Parshani, Roni and Paul, Gerald and Stanley, H. Eugene and Havlin, Shlomo},
  journal={Nature},
  volume={464},
  number={7291},
  pages={1025--1028},
  year={2010},
  publisher={Nature Publishing Group},
  doi={10.1038/nature08932}
}

@article{Havlin2009CatastrophicCO,
  title={Catastrophic cascade of failures in interdependent networks},
  author={Shlomo Havlin and Nuno A. M. Ara{\'u}jo and Sergey V. Buldyrev and et al.},
  journal={Nature},
  year={2009},
  volume={464},
  pages={1025-1028}
}

@article{Gao2011NetworksFF,
  title={Networks formed from interdependent networks},
  author={Jianxi Gao and Sergey V. Buldyrev and H. Eugene Stanley and et al.},
  journal={Nature Physics},
  year={2011},
  volume={8},
  pages={40 - 48}
}

@article{gao2016universal,
  title={Universal resilience patterns in complex networks},
  author={Gao, Jianxi and Barzel, Baruch and Barab{\'a}si, Albert-L{\'a}szl{\'o}},
  journal={Nature},
  volume={530},
  number={7590},
  pages={307--312},
  year={2016},
  publisher={Nature Publishing Group}
}

@article{Parshani2010CriticalEO,
  title={Critical effect of dependency groups on the function of networks},
  author={Roni Parshani and Sergey V. Buldyrev and Shlomo Havlin},
  journal={Proceedings of the National Academy of Sciences},
  year={2010},
  volume={108},
  pages={1007 - 1010}
}

@article{danziger2022recovery,
  title={Recovery coupling in multilayer networks},
  author={Danziger, Michael M and Barab{\'a}si, Albert-L{\'a}szl{\'o}},
  journal={Nature Communications},
  volume={13},
  number={1},
  pages={955},
  year={2022},
  publisher={Nature Publishing Group}
}

@article{mahabadi2021network,
  title={Network properties for robust multilayer infrastructure systems: A percolation theory review},
  author={Mahabadi, Zahra and Varga, Liz and Dolan, Tom},
  journal={IEEE Access},
  volume={9},
  pages={135755--135773},
  year={2021},
  publisher={IEEE}
}

@article{ferrazdearruda2024contagion,
  title={Contagion dynamics on higher-order networks},
  author={Ferraz de Arruda, Guilherme and Aleta, Alberto and et al.},
  journal={Nature Reviews Physics},
  volume={6},
  number={8},
  pages={468--482},
  year={2024},
  publisher={Nature Publishing Group UK London}
}

@article{holling1973resilience,
  author = {Holling, Crawford S.},
  title = {Resilience and Stability of Ecological Systems},
  journal = {Annual Review of Ecology, Evolution, and Systematics},
  volume = {4},
  pages = {1--23},
  year = {1973}
}

@article{martin2015regional,
  author = {Martin, Ron and Sunley, Peter},
  title = {On the Notion of Regional Economic Resilience: Conceptualization and Explanation},
  journal = {Journal of Economic Geography},
  volume = {15},
  number = {1},
  pages = {1--42},
  year = {2015}
}

@article{proag2014concept,
  author = {Proag, Virendra},
  title = {The Concept of Vulnerability and Resilience},
  journal = {Procedia Economics and Finance},
  volume = {18},
  pages = {369--376},
  year = {2014}
}

@article{wied2020conceptualizing,
  author = {Wied, M. and Oehmen, J. and Welo, T.},
  title = {Conceptualizing Resilience in Engineering Systems: An Analysis of the Literature},
  journal = {Systems Engineering},
  volume = {23},
  pages = {3--13},
  year = {2020}
}

@article{bergstrom2015rationale,
  author = {Bergstr{\"o}m, Johan and van Winsen, Roel and Henriqson, Eder},
  title = {On the Rationale of Resilience in the Domain of Safety: A Literature Review},
  journal = {Reliability Engineering \& System Safety},
  volume = {141},
  pages = {131--141},
  year = {2015},
  issn = {0951-8320}
}

@article{haimes2008homeland,
  title={Homeland security preparedness: Balancing protection with resilience in emergent systems},
  author={Haimes, Yacov Y and Crowther, Kenneth and Horowitz, Barry M},
  journal={Systems Engineering},
  volume={11},
  number={4},
  pages={287--308},
  year={2008}
}

@article{XU2024tsosra,
title = {TSoSRA: A task-oriented resilience assessment framework for system-of-systems},
journal = {Reliability Engineering \& System Safety},
volume = {248},
pages = {110186},
year = {2024},
issn = {0951-8320},
author = {Renjie Xu and Jiahao Liu and Jichao Li and Kewei Yang and et al.}
}

@article{Dui2023ImportanceMR,
  title={Importance measure-based resilience management: Review, methodology and perspectives on maintenance},
  author={Hongyan Dui and Meng Liu and Jiaying Song and et al.},
  journal={Reliability Engineering \& System Safety},
  year={2023},
  volume={237},
  pages={109383}
}

@article{akyildiz2005uans,
  author    = {Akyildiz, Ian F. and Pompili, Dario and Melodia, Tommaso},
  title     = {Underwater acoustic sensor networks: research challenges},
  journal   = {Ad Hoc Networks},
  volume    = {3},
  number    = {3},
  pages     = {257--279},
  year      = {2005},
  publisher = {Elsevier}
}

@inproceedings{sarker2024low,
  title={Low latency and resilient industrial network with next generation SDN and PLC virtualization},
  author={Sarker, Taposh Kumer and Tosh, Deepak K},
  booktitle={IEEE Military Communications Conference},
  pages={1040--1045},
  year={2024},
  organization={IEEE}
}

@article{farhan2025rdg,
  title={RDG-TE: Link reliability-aware DRL-GNN-based traffic engineering in SDN},
  author={Farhan, Muhammad and Shah, Nadir and Wang, Lei and et al.},
  journal={Expert systems with applications},
  volume={265},
  pages={125963},
  year={2025},
  publisher={Elsevier}
}

@article{pedroso2025survey,
  title={A Survey on Faults and Vulnerabilities in Network Topological Connectivity: Logical and Physical Perspectives},
  author={Pedroso, Carlos and Santos, Aldri},
  journal={ACM Computing Surveys},
  year={2025},
  publisher={ACM New York, NY}
}

@article{alhidaifi2024survey,
  title={A survey on cyber resilience: Key strategies, research challenges, and future directions},
  author={Alhidaifi, Saleh Mohamed and Asghar, Muhammad Rizwan and Ansari, Imran Shafique},
  journal={ACM computing surveys},
  volume={56},
  number={8},
  pages={1--48},
  year={2024},
  publisher={ACM New York, NY}
}

@article{kim2022survey,
  title={A survey on network security for cyber--physical systems: From threats to resilient design},
  author={Kim, Sangjun and Park, Kyung-Joon and Lu, Chenyang},
  journal={IEEE Communications Surveys \& Tutorials},
  volume={24},
  number={3},
  pages={1534--1573},
  year={2022},
  publisher={IEEE}
}

@article{chen2022low,
  author={Chen, Quan and Yang, Lei and Guo, Jianming and et al.},
  title={LEO Mega-Constellation Network: Networking Technologies and State of the Art},
  journal={Journal on Communications},
  year={2022},
  volume={43},
  number={5},
  pages={177--189}
}

@inproceedings{kang2025deepdive,
author = {Kang, Eunju and Ramanathan, Alagappan and Jyothi, Sangeetha Abdu},
title = {A Deep Dive into the Impact of Solar Storms on LEO Satellite Networks},
year = {2025},
isbn = {9798400720901},
publisher = {Association for Computing Machinery},
address = {New York, NY, USA},
url = {https://doi.org/10.1145/3748749.3749094},
doi = {10.1145/3748749.3749094},
booktitle = {Proceedings of the 2025 3rd Workshop on LEO Networking and Communication, Mobicom 2025},
pages = {74–81},
numpages = {8},
series = {LEO-NET '25}
}

@article{oneil2025enhancingcritical,
    author = {O'Neil, R. and Diallo, C. and Khatab, A. and et al.},
    title = {Enhancing Critical Network Infrastructure Resilience Through Optimal Post-Disruption Maintenance and Routing Decisions},
    journal = {Reliability Engineering and System Safety},
    volume = {257},
    number = {110717},
    year = {2025},
    publisher = {Elsevier}
}

@article{berger2021survey,
  title={A survey on resilience in the iot: Taxonomy, classification, and discussion of resilience mechanisms},
  author={Berger, Christian and Eichhammer, Philipp and Reiser, Hans P and et al.},
  journal={ACM Computing Surveys},
  volume={54},
  number={7},
  pages={1--39},
  year={2021},
  publisher={ACM New York, NY, USA}
}

@article{cassottana2023resilience,
  title={Resilience analysis of cyber-physical systems: A review of models and methods},
  author={Cassottana, Beatrice and Roomi, Muhammad M and Mashima, Daisuke and et al.},
  journal={Risk Analysis},
  volume={43},
  number={11},
  pages={2359--2379},
  year={2023},
  publisher={Wiley Online Library}
}

@article{zhu2025small,
  title={A small set of critical hyper-motifs governs heterogeneous flow-weighted network resilience},
  author={Zhu, Tianlei and Yang, Xin and Ma, Zhiao and et al.},
  journal={Nature Communications},
  volume={16},
  number={1},
  pages={7809},
  year={2025},
  publisher={Nature Publishing Group UK London}
}

@article{duan2025resilience,
  title={Resilience prediction and tipping point control of multilayer ecological networks based on dimensionality reduction method},
  author={Duan, Dongli and Zhao, Xingjie and Cai, Zhiqiang and et al.},
  journal={Chaos, Solitons \& Fractals},
  volume={191},
  pages={115914},
  year={2025},
  publisher={Elsevier}
}

@article{wu2024collapse,
  title={Collapse process prediction of mutualistic dynamical networks with k-core and dimension reduction method},
  author={Wu, Chengxing and Duan, Dongli},
  journal={Chaos, Solitons \& Fractals},
  volume={180},
  pages={114489},
  year={2024},
  publisher={Elsevier}
}

@article{kaiser2024network,
  title={Network isolators inhibit failure spreading in complex networks},
  author={Kaiser, F and Latora, V and Witthaut, D},
  journal={Nature Communications},
  volume={12},
  number={1},
  pages={3143},
  year={2024}
}

@article{zhu2025can,
  title={Can a bottleneck capacity enhancement policy postpone infrastructure network collapse?},
  author={Zhu, Tianlei and Yang, Xin and Ma, Zhiao and et al.},
  journal={Transport Policy},
  volume={163},
  pages={1--13},
  year={2025},
  publisher={Elsevier}
}

@article{qiu2023self,
  title={A self-adaptive robustness optimization method with evolutionary multi-agent for IoT topology},
  author={Qiu, Tie and Yang, Xinwei and Chen, Ning and et al.},
  journal={IEEE/ACM Transactions on Networking},
  volume={32},
  number={2},
  pages={1346--1361},
  year={2023},
  publisher={IEEE}
}

@article{chen2024daimo,
  title={Daimo: motif density enhances topology robustness for highly dynamic scale-free iot},
  author={Chen, Ning and Qiu, Tie and Si, Weisheng and Wu, Dapeng Oliver},
  journal={IEEE Transactions on Mobile Computing},
  year={2024},
  publisher={IEEE}
}

@article{callaway2000robustness,
  title={Network robustness and fragility: Percolation on random graphs},
  author={Callaway, Duncan S. and Newman, M. E. J. and Strogatz, Steven H. and Watts, Duncan J.},
  journal={Physical Review Letters}, volume={85}, number={25}, pages={5468--5471}, year={2000}
}

@article{cohen2001breakdown,
  title={Breakdown of the Internet under intentional attack},
  author={Cohen, Reuven and Erez, Keren and ben-Avraham, Daniel and Havlin, Shlomo},
  journal={Physical Review Letters}, volume={86}, number={16}, pages={3682--3685}, year={2001}
}

@article{achlioptas2009explosive,
  title={Explosive percolation in random networks},
  author={Achlioptas, Dimitris and D'Souza, Raissa M. and Spencer, Joel},
  journal={Science}, volume={323}, number={5920}, pages={1453--1455}, year={2009}
}

@article{kitsak2010identification,
  title={Identification of influential spreaders in complex networks},
  author={Kitsak, Maksim and Gallos, Lazaros K. and Havlin, Shlomo and Liljeros, Fredrik and Muchnik, Lev and Stanley, H. Eugene and Makse, Hern{\'a}n A.},
  journal={Nature Physics}, volume={6}, number={11}, pages={888--893}, year={2010}
}

@article{helbing2013globally,
  title={Globally networked risks and how to respond},
  author={Helbing, Dirk},
  journal={Nature}, volume={497}, number={7447}, pages={51--55}, year={2013}
}

@book{newman2010networks,
  title={Networks: An Introduction},
  author={Newman, M. E. J.},
  publisher={Oxford University Press}, address={Oxford, U.K.}, year={2010}
}

@article{netbreaker2026,
  title={{NetBreaker}: Benchmarking Network Dismantling across Objectives, Scales, and Topologies},
  author={Yang, Wenchuan and Bu, Tianci and Tan, Suoyi and Fan, Changjun and Zhou, Zijie and Lu, Xin},
  journal={arXiv preprint arXiv:2611.19298}, year={2026}
}

@article{targetaware2026,
  title={Target-Aware Network Dismantling with Limited Budgets},
  author={Song, Jiaqi and Jing, Fei and Huang, Wenjie and Qi, Xingqin and Zhang, Qingpeng},
  journal={arXiv preprint arXiv:2610.07926}, year={2026}
}

@article{criticality2026,
  title={Criticality and universality in network dismantling},
  author={Cirigliano, Lorenzo and Castellano, Claudio and Kim, Minsuk and Radicchi, Filippo and Sun, Hanlin},
  journal={arXiv preprint arXiv:2608.27613}, year={2026}
}

@article{autosnd2026,
  title={{AutoSND}: From Execution Evidence to Structural Policies for Automated Network Dismantling Heuristic Discovery},
  author={Hu, Zhijing and Fan, Changjun and Deng, Yufan and Cao, Zhiguang},
  journal={arXiv preprint arXiv:2608.03653}, year={2026}
}

@article{hypergraphdismantling2026,
  title={Identifying vulnerable nodes for hypergraph dismantling via higher-order competition dynamics},
  author={Hao, Yajing and Liu, Longzhao and Wang, Xin and Han, Zhihao and Zheng, Hongwei and Tang, Shaoting},
  journal={arXiv preprint arXiv:2606.23289}, year={2026}
}

@article{directed2025,
  title={Optimal dismantling of directed networks},
  author={Liu, Xueming and Hu, Jiawen and Wang, Yumei and Liu, Yang-Yu and Zhang, Hai-Tao},
  journal={arXiv preprint arXiv:2512.11416}, year={2025}
}

@article{causalembed2025,
  title={Importance Ranking in Complex Networks via Influence-aware Causal Node Embedding},
  author={Gao, Jiahui and Zhou, Kuang and Zhu, Yuchen and Wu, Keyu},
  journal={arXiv preprint arXiv:2511.01228}, year={2025}
}

@article{nohandcraft2025,
  title={Learning Network Dismantling Without Handcrafted Inputs},
  author={Tian, Haozhe and Ferraro, Pietro and Shorten, Robert and Jalili, Mahdi and Hamedmoghadam, Homayoun},
  journal={arXiv preprint arXiv:2508.00706}, year={2025}
}

@article{physicaldamage2024,
  title={Network dismantling by physical damage},
  author={Blagojevi{\'c}, Luka and Bonamassa, Ivan and P{\'o}sfai, M{\'a}rton},
  journal={arXiv preprint arXiv:2412.09524}, year={2024}
}

@article{esnd2024,
  title={{ESND}: An Embedding-based Framework for Signed Network Dismantling},
  author={Xie, Chenwei and Liu, Chuang and Li, Cong and Zhan, Xiu-Xiu and Li, Xiang},
  journal={arXiv preprint arXiv:2406.08899}, year={2024}
}

@article{embeddismantle2022,
  title={Embedding-aided network dismantling},
  author={Osat, Saeed and Papadopoulos, Fragkiskos and Teixeira, Andreia Sofia and Radicchi, Filippo},
  journal={arXiv preprint arXiv:2208.01087}, year={2022}
}

@article{smallnetworks2025,
  title={Robustness of `small' networks},
  author={Jiang, Jessica and Zhuang, Allison C. and Holme, Petter},
  journal={arXiv preprint arXiv:2509.23670}, year={2025}
}

@article{higherordercascade2025,
  title={Robustness of One-to-Many Interdependent Higher-order Networks Against Cascading Failures},
  author={Qian, Cheng and Zhao, Dandan and Zhang, Bo},
  journal={arXiv preprint arXiv:2509.23716}, year={2025}
}

@article{quantumperc2026,
  title={Quantum percolation theory for dynamic propagation connectivity of transport networks},
  author={Xu, Junxiang and Niu, Chence and Nair, Divya Jayakumar},
  journal={arXiv preprint arXiv:2606.30218}, year={2026}
}

@article{digitaltwinleo2026,
  title={Digital Twin Satellite Networks: A Paradigm for Intelligent, Efficient, and Resilient Operations},
  author={Alhassan, Mustafa and Hu, Peng},
  journal={arXiv preprint arXiv:2608.12865}, year={2026}
}

@article{swarmtopo2025,
  title={Criticality-Based Dynamic Topology Optimization for Enhancing Aerial-Marine Swarm Resilience},
  author={Huang, Ruiyang and Wang, Haocheng and Shen, Yixuan},
  journal={arXiv preprint arXiv:2508.00688}, year={2025}
}

@inproceedings{egnd2020,
  title={Ensemble approach for generalized network dismantling},
  author={Ren, Xiao-Long and Antulov-Fantulin, Nino},
  booktitle={Complex Networks and Their Applications VIII},
  series={Studies in Computational Intelligence}, pages={783--793},
  publisher={Springer}, year={2020}
}

@inproceedings{coregdm2023,
  title={{CoreGDM}: Geometric Deep Learning Network Decycling and Dismantling},
  author={Grassia, Marco and Mangioni, Giuseppe},
  booktitle={Complex Networks XIV (CompleNet 2023)},
  series={Springer Proceedings in Complexity}, publisher={Springer}, year={2023}
}

\end{document}